\documentclass[fleqn,usenatbib]{mnras}

\usepackage{newtxtext,newtxmath}

\usepackage[T1]{fontenc}

\DeclareRobustCommand{\VAN}[3]{#2}
\let\VANthebibliography\thebibliography
\def\thebibliography{\DeclareRobustCommand{\VAN}[3]{##3}\VANthebibliography}

\usepackage{graphicx}	
\usepackage{amsmath}	
\usepackage{longtable}

\title[Validation of Sub-Saturn Exoplanet TOI-181b]{VaTEST I: Validation of Sub-Saturn Exoplanet TOI-181b in Narrow Orbit from its Host Star}

\author[Priyashkumar Mistry et al.]{Priyashkumar Mistry,$^{1}$\thanks{E-mail: i17ph033@phy.svnit.ac.in}
Kamlesh Pathak,$^{1}$
Georgios Lekkas,$^{2}$
Aniket Prasad,$^{3}$
Surendra Bhattarai,$^{4}$
\newauthor
Mousam Maity,$^{5}$
Charles A. Beichman,$^{6}$
David R. Ciardi,$^{6}$
Phil Evans$,^{7}$
Allyson Bieryla,$^{8}$
\newauthor
Jason D. Eastman,$^{8}$
David W. Latham,$^{8}$
Gilbert A. Esquerdo$^{8}$
and Jennifer P. Lucero$^{9}$
\\
\\
$^{1}$Department of Physics, Sardar Vallabhbhai National Institute of Technology, Surat-395007, Gujarat, India\\
$^{2}$Department of Physics, University of Ioannina, Ioannina-45110, Greece\\
$^{3}$Department of Physics, National Institute of Technology, Agartala-799046, Tripura, India\\
$^{4}$Department of Physical Sciences, Indian Institute of Science Education and Research Kolkata, Mohanpur-741246, West Bengal, India\\
$^{5}$Department of Physics, Presidency University, Kolkata-700073, West Bengal, India\\
$^{6}$NASA Exoplanet Science Institute - Caltech/IPAC Pasadena, CA 91350 USA.\\
$^{7}$El Sauce Observatory, Coquimbo Province, Chile.\\
$^{8}$Center for Astrophysics, Harvard \& Smithsonian, 60 Garden St, Cambridge, MA 02138, USA\\
$^{9}$Institute of Astrophysics, Pontificia Universidad Católica de Chile, Santiago-8331150, Región Metropolitana, Chile
}

\date{Accepted XXX. Received YYY; in original form ZZZ}

\pubyear{2023}

\begin{document}
\label{firstpage}
\pagerange{\pageref{firstpage}--\pageref{lastpage}}
\maketitle

\begin{abstract}
We present here a validation of sub-Saturn exoplanet TOI-181b orbiting a K spectral type star TOI-181 (Mass: 0.822 $\pm$ 0.04 M$_{\sun}$, Radius: 0.745 $\pm$ 0.02 R$_{\sun}$, Temperature: 4994 $\pm$ 50 K) as a part of \textbf{Va}lidation of \textbf{T}ransiting \textbf{E}xoplanets using \textbf{S}tatistical \textbf{T}ools (VaTEST) project. TOI-181b is a planet with radius 6.95 $\pm$ 0.08 R$_{\earth}$, mass 46.16 $\pm$ 2.71 M$_{\earth}$, orbiting in a slightly eccentric orbit with eccentricity 0.15 $\pm$ 0.06 and semi-major axis of 0.054 $\pm$ 0.004 AU, with an orbital period of 4.5320 $\pm$ 0.000002 days. The transit photometry data was collected using Transiting Exoplanet Survey Satellite (TESS) and spectroscopic data for radial velocity analysis was collected using The European Southern Observatory's (ESO) High Accuracy Radial Velocity Planet Searcher (HARPS) telescope. Based on the radial velocity best-fit model we measured RV semi-amplitude to be 20.56 $\pm$ 2.37 m s$^{-1}$. Additionally, we used \texttt{VESPA} and \texttt{TRICERATOPS} to compute the False Positive Probability (FPP), and the findings were FPP values of $1.68\times10^{-14}$ and $3.81\times10^{-04}$, respectively, which are significantly lower than the 1\% threshold. The finding of TOI-181b is significant in the perspective of future work on the formation and migration history of analogous planetary systems since warm sub-Saturns are uncommon in the known sample of exoplanets.
\end{abstract}

\begin{keywords}
individual: TOI-181, technique: photometry, spectroscopic, radial velocities, methods: statistical
\end{keywords}


\section{Introduction}
\label{sec:1}
Up to the time of writing a total number of 5054 exoplanets have been discovered and confirmed \footnote{\url{https://exoplanetarchive.ipac.caltech.edu/}}. For the coming years, this amount is expected to experience an even larger growth with many near and far future missions expected to be launched. Many of these discovered extra-solar planets are different from the planets of our Solar System, imposing a wide variety of different exoplanets that can exist. The very first exoplanet discovery happened 30 years ago, in 1992, by the discoverers Aleksander Wolszczan and Dale Frail \citep{1992Natur.355..145W}, who found two exoplanets orbiting a pulsar.

Astronomers have discovered and classified many types of exoplanets. In addition to super-Earths, which are terrestrial planets composed of rocky materials and larger than our own, there are also hot-Jupiters. These planets are comparable in size to or larger than Jupiter, and they orbit their host star very closely. Sub-Saturns or super-Neptunes is a further classification. This category of exoplanets is under-explored and incredibly interesting to investigate due to their unique properties. The exoplanets discovered in this category have a narrow range of radii (2-8 R$_{\earth}$) but larger masses (20-100 M$_{\earth}$). Sub-Saturns have very little correlation between their masses and radii and strong relation with the metallicity of their host stars \citep{2017AJ....153..142P}. According to theory, hot Jupiters and sub-Saturns cannot originate within the ice line because their gaseous envelopes would be torn apart by stellar wind, leaving only their low-mass rocky cores. According to the concept of planetary migration, this sort of planet forms outside the ice line and migrates toward its host star through an exchange of angular momentum \citep{lissauer2007formation}.

The launch of space missions like Kepler \citep{borucki2010kepler}, K2 \citep{howell2014k2}, and TESS \citep{2015AAS...22520201R} have provided us with valuable data which resulted in the discovery of a huge number of exoplanets. However, a pretty common thing, space missions can suffer from, is the detection of the so-called False Positive events. False positives or false detection are cases in which a transit signal from the target is not caused by a true exoplanet transit but rather by any other background source or eclipsing binaries. False detection in transit space missions can be filtered out by statistical validation of transit candidates. Hence, concentrated efforts have been made to develop efficient statistical validation tools that can be employed for various space missions. \texttt{BLENDER} \citep{2005ApJ...619..558T} was one of the first statistical tools utilised for validating exoplanet candidates from Kepler. It used $\chi^2$ statistics on synthetically generated light curves based on planetary and blended eclipsing binary models combined with multi-colour photometry calculation to rule out the possibilities of false positive scenarios. \texttt{BLENDER}’s approach to testing for false positive astrophysical scenarios provided a framework for exoplanet validation despite its suffering from long computation times owing to the simulation of false positive scenarios. A newer validation framework by the name of \texttt{VESPA} \citep{2012ApJ...761....6M} greatly reduced computation time by employing a simpler trapezoidal model in place of physical transit models, as was the case with \texttt{BLENDER}. Using the TRILEGAL \citep{2005A&A...436..895G} model of the galaxy, \texttt{VESPA} simulates a population of stars similar in properties to the target star. The population is then used to produce a prior distribution for each astrophysical scenario using the trapezoidal model. It used an MCMC sampling routine to fit the Kepler light curve and produced a false positive probability based on the fit. Both \texttt{VESPA} and \texttt{BLENDER} have the capability to account for follow-up high contrast imaging. The framework was widely used to statistically validate exoplanets from Kepler as well as TESS. \texttt{TRICERATOPS} \citep{2020ascl.soft02004G} was developed specifically to utilise the unique features and requirements of the TESS mission. With a lower resolution than previous such missions, there is an even greater necessity to account for multiple star systems and scenarios like diluted transits. \texttt{VESPA} and \texttt{TRICERATOPS} have been used to successfully validate planet candidates for the Kepler \citep{2016ApJ...822...86M}, K2 \citep{2017AJ....154..207D,2021AAS...23810804B,2022AJ....163..244C} and TESS \citep{2021AJ....161...24G} missions.

Other tools such as \texttt{PASTIS} \citep{2014MNRAS.441..983D,2015MNRAS.451.2337S} have also been alternatively used to validate exoplanet candidates. \texttt{PASTIS} is unique in the sense that it provides the capability to combine radial velocity observation with light curve analysis for validation. However, this does make \texttt{PASTIS} a more robust framework to validate transiting exoplanets at the cost of being computationally expensive. It also utilises a realistic model of galactic extinction on top of the TRILEGAL galactic model to better account for the possible blends, which crucially depends on the frequency of background star extinction in the line of sight. Using two of these statistical tools (chosen to independently verify the result and to utilise the merits of statistical tools, discussed in section \ref{sec:4}) along with the TESS photometry data and radial velocity follow-up data from HARPS \citep{2000SPIE.4008..582P}, we present here the evidence for the existence of a sub-Saturn planet orbiting TOI-181 as observed by TESS in sectors 2 and 29.

The statistical validation used in this current study is part of the newly launched project, \textbf{Va}lidation of \textbf{T}ransiting \textbf{E}xoplanets using \textbf{S}tatistical \textbf{T}ools (VaTEST)\footnote{\url{https://sites.google.com/view/project-vatest/home}}. The aim of this project is to validate exoplanet signals using statistical tools and probabilistic algorithms, in combination with other packages mentioned later in this study (\texttt{Lightkurve} \citep{2018ascl.soft12013L}, \texttt{TLS} \citep{2019A&A...623A..39H}, and \texttt{Juliet} \citep{2019MNRAS.490.2262E} for photometric and spectroscopic data analysis. Future work of this project will be aimed at bulk validation of unconfirmed exoplanets using a variety of statistical validation, false alarm and false positive diagnostics tests and photometric and spectroscopic data analysis to calculate the exoplanet properties.

The rest of this paper is organized as follows. In Section \ref{sec:2}, we present the observational transit data from TESS, ground based transit follow-up, high resolution imaging observation and radial velocity data from HARPS. In section and \ref{sec:3}, we focus our attention to the host star and characterise the star-planet system using a combined model of transit photometry and RVs. In section In Section \ref{sec:4}, we discussed prominent features of confirmed exoplanet TOI-181b. Section \ref{sec:5} refers to the summary of this paper and the conclusion.

\section{Data Reduction}
\label{sec:2}
\subsection{Transit Photometry}
\label{sec:2.1}
\begin{figure*}
\begin{center}
\includegraphics[scale=0.65]{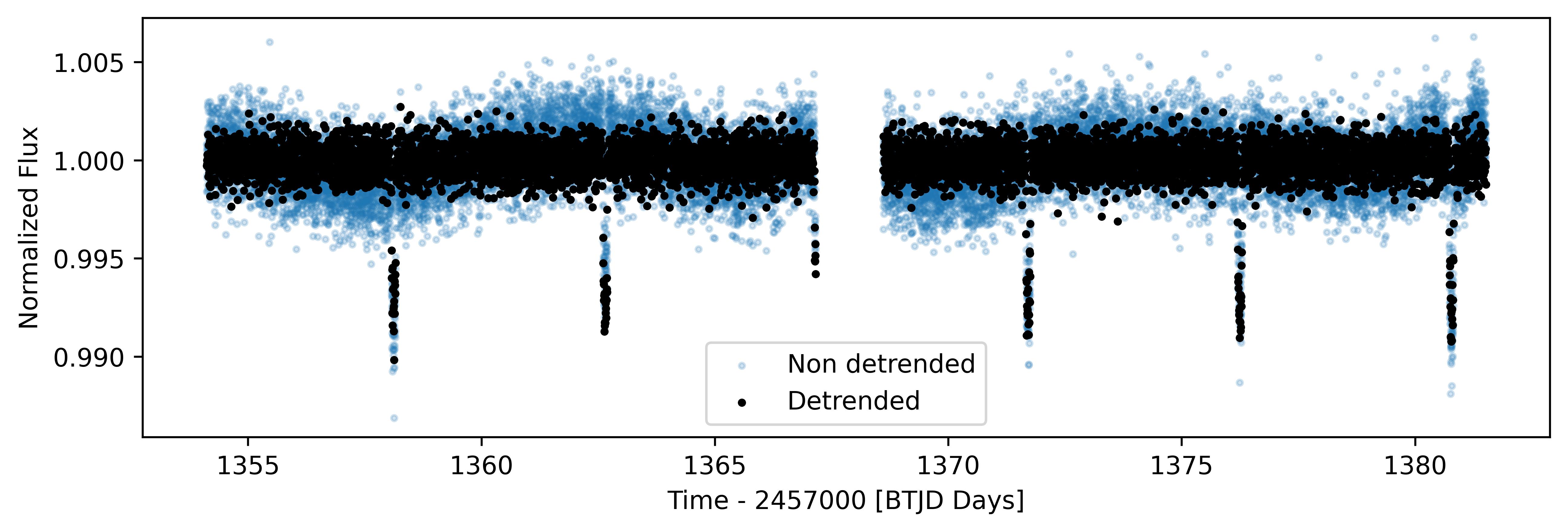}
\includegraphics[scale=0.65]{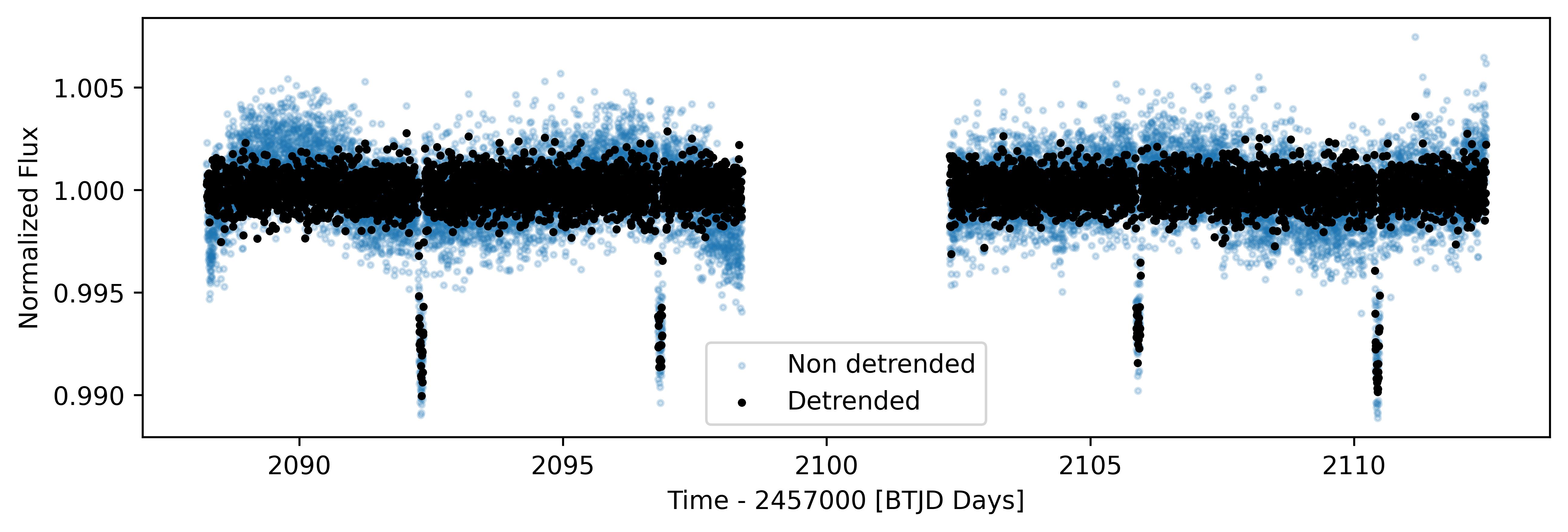}
\end{center}
\caption{Transit light curve of sector 02 (upper panel) and sector 29 (lower panel). Blue dots represent the non-de-trended light curve and black dots are for de-trended light curve. De-trended light curve is binned so that difference can be clearly visible.}
\label{fig:01}
\end{figure*}

TOI-181 (TIC 76923707) can be found at the coordinates RA 23:28:41.41 and Dec -34:29:29.08, and it was spotted by the Transiting Exoplanet Search Satellite (TESS) \citep{2015AAS...22520201R} in sector 2 on August 23rd, 2018 and sector 29 on August 26th, 2020. In sector 2, photometric data for TOI-181 was collected with a 2-minute cadence, whereas data for sector 29 was collected with a 2-minute as well as a 20-second cadence. These pixel data were processed by the TESS Science Processing Operation Centre (TESS-SPOC) pipeline \citep{2016SPIE.9913E..3EJ}, and made available the target pixel files (TPF) and light curve files cleaned from instrumental systematics. The Transiting Planet Search (TPS) module of SPOC was responsible for the independent recovery of transit signals for each sector. With a signal-to-noise ratio (SNR) of 86.57, the SPOC pipeline was successful in recovering a transiting signal for TOI-181. The transit signal for this specific target is shown in Fig. \ref{fig:01}.

With the help of the \texttt{Lightkurve} Python package \citep{2018ascl.soft12013L}, we were able to download these signals from the Mikulski Archive for Space Telescopes (MAST)\footnote{\url{https://archive.stsci.edu/}}. This light curve exhibited stellar variability, which was eliminated by de-trending the light curve using the \texttt{flatten} function of \texttt{Lightkurve}. In order to ensure that the in-transit portion of a signal was not lost during the de-trending process, we used a mask that was 0.12-days long and had a period of 4.53 days, starting from the epoch time 2458358.12. Fig. \ref{fig:01} illustrates the difference that can be seen between the signal after it has been de-trended and the actual signal.

We determined the duration of the transit event, the time of the epoch, and the transit depth by applying a Transit Least Squares (\texttt{TLS}) \citep{2019A&A...623A..39H} model to the light curve after it had been cleaned up and de-trended. The values that are reported by SPOC on the ExoFOP-TESS\footnote{\url{https://exofop.ipac.caltech.edu/tess/}} website are identical to the values that are calculated by \texttt{TLS}.

\subsection{Follow-up Transit Photometry}
\label{sec:2.2}
\begin{figure*}
\begin{center}
\includegraphics[scale=0.6]{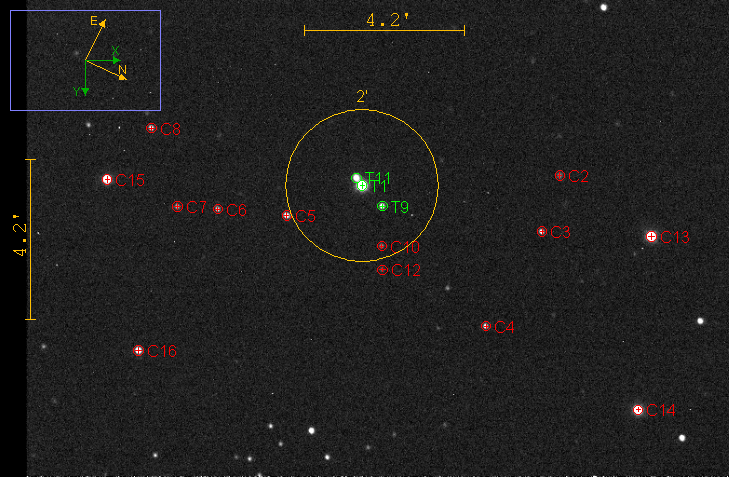}\\
\includegraphics[scale=0.5]{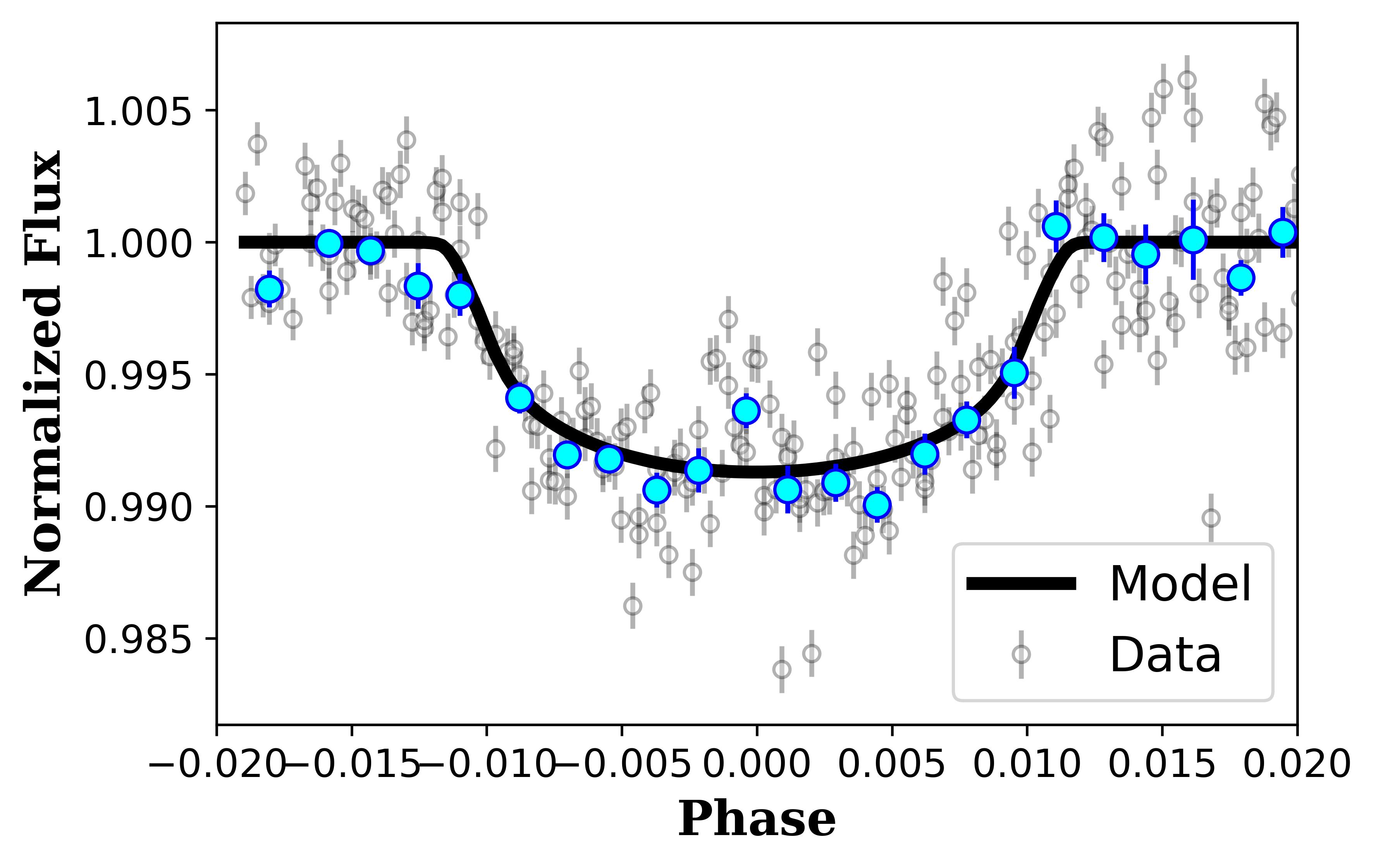}
\end{center}
\caption{Upper Panel: El Sauce Observatory field of view. Yellow circle shows 2$\arcmin$ region from our target. Green circle marked with T1 is TOI-181, T9 and T11 are stars that lie within 2$\arcmin$ region. Stars marked with red circles are comparison stars. Lower Panel: Transit light curve of target TOI-181, relative flux was calculated using comparison stars. Photometry data was collected using telescope equipped with STT 1603-3 CCD camera at El Sauce Observatory. Blue filled circles shows the binned data.}
\label{fig:02}
\end{figure*}

We obtained ground-based time-series photometry in order to precisely collect the transit features and to comprehend and confirm the nature of the transit detected by TESS. This follow-up observation was conducted as part of the TESS Follow-up Observing Program (TFOP; \citet{2018AAS...23143908C}). We used the TESS Transit Finder, a customized version of the Tapir software package \citep{2013ascl.soft06007J} to schedule our observations. We observed a full transit in Johnson-Cousins R-band on UT20181108 using the Evans 0.36m telescope at El Sauce Observatory in Coquimbo Province, Chile. The telescope was equipped with an STT 1603-3 CCD camera with 1536 $\times$ 1024 pixels binned 2 $\times$ 2 in-camera resulting in an image scale of 1.47" /pixel. The photometric data was obtained from 10 $\times$ 70 seconds followed by 254 $\times$ 60 seconds exposures, after standard calibration, using a circular 7.4" aperture in AstroImageJ \citep{2016arXiv160102622C}. Fig. \ref{fig:02} shows the field of view of El Sauce Observatory along with the transit light curve obtained by calculating the relative flux of target star (marked with green circle and T1) with respect to other comparison stars.

\subsection{High Angular Resolution Imaging}
\label{sec:2.3}
High resolution imaging is an effective tool for reducing the possibility of blended background objects. We used the archival, high-angular-resolution AO image data collected by the Keck II/NIRC2 camera. Additionally, ExoFOP makes this publicly accessible (submitted by David Ciardi). The Keck AO data was collected on December 06, 2018. The pixel scale for NIRC2 is 9.942 mas per pixel. 

\begin{figure}
\begin{center}
\includegraphics[scale=0.12]{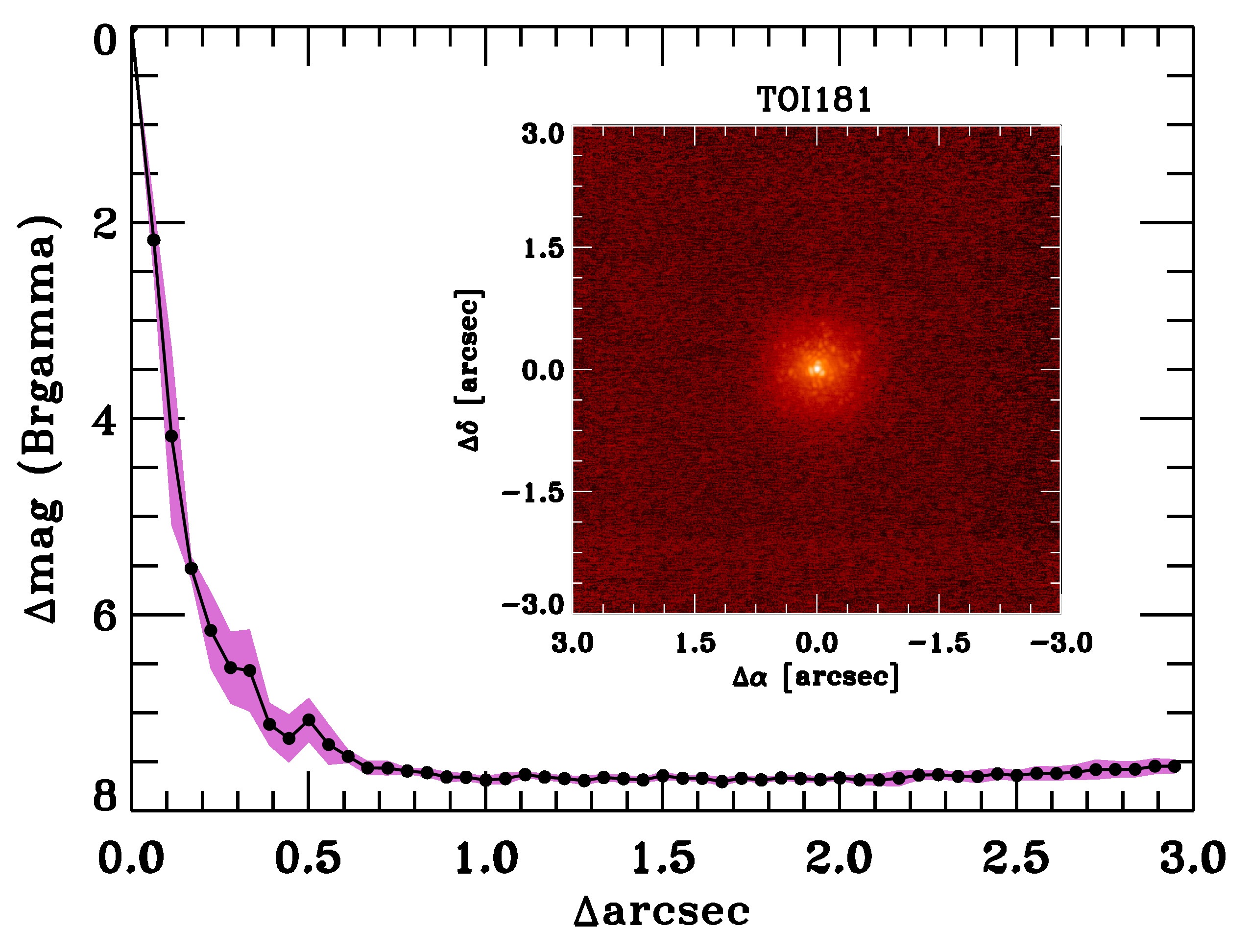}
\end{center}
\caption{High resolution image of TOI-181 with 5$\sigma$ contrast curve (uploaded on ExoFOP by David Ciardi). Pink shaded region describes the uncertainties in the contrast sensitivity estimated by measuring RMS dispersion. Estimated sensitivity of companion is $\approx$7 $\Delta$mag at 0.5". Curve is saturating $\approx$8 $\Delta$mag, so the point source that has contrast greater than this value will remain non-detected.}
\label{fig:highreso}
\end{figure}

Fig. \ref{fig:highreso} displays the observed image. The estimated point spread function of the source is 0.0556". The star looks to be single and doesn't have any very close neighbours within a few arcseconds. At a distance of 0.5", the estimated sensitivity to the companion is $\approx$7 $\Delta$mag. So eventually, it rules out the possibility of any background blended source affecting the light curve of our target.

\subsection{Radial Velocity Observations}
\label{sec:2.4}
The European Southern Observatory's (ESO) High Accuracy Radial Velocity Planet Searcher (HARPS) \citep{2000SPIE.4008..582P} telescope collected spectral data in order to calculate the radial velocity (RV) of the host star. HARPS is a 3.6-meter telescope that is based at ESO's La Silla Observatory. HARPS can achieve radial velocity precision on the order of 1 m/s over very long timescales. To avoid spectral drift due to temperature and air pressure, spectrograph was fed with two fibres contained in a vacuum chamber. One of the fibres collects starlight, while the other simultaneously measures the Th-Ar reference spectrum. Both fibres have a 1" aperture in the sky. We were able to recover 23 spectral observations between the dates of June 25th, 2019 and December 1st, 2020. The spectral range of these observations was 378.2-691.3 nm, and the spectral resolution was 115000. We made use of a pipeline that had been developed by Nora Eisner \citep{EisnerIdentifying} in order to obtain the RV values from these spectra. The RV data had been extracted and are presented in Table \ref{tab:01}.

\begin{table}
\caption{Radial velocity data as extracted from the spectral data collected by HARPS. Based on data obtained from the ESO Science Archive Facility with DOI: \url{https://doi.org/10.18727/archive/33}}
\label{tab:01}
\begin{center}
\begin{tabular}{c c c}
\hline
BJD Time & RV (m s$^{-1}$) & RV Uncertainty (m s$^{-1}$) \\
\hline
\hline
2458659.85568	&	-4766.142504	&	8.435472	\\
2458660.85346	&	-4745.218259	&	6.531391	\\
2458670.77969	&	-4756.403730	&	4.848487	\\
2458672.86062	&	-4752.448405	&	6.362948	\\
2458673.83568	&	-4728.360590	&	3.715174	\\
2458674.87044	&	-4739.506311	&	5.139311	\\
2458675.81618	&	-4755.954379	&	4.945969	\\
2458683.86426	&	-4750.973759	&	1.906334	\\
2458683.94273	&	-4747.586764	&	6.552498	\\
2458684.79363	&	-4772.551952	&	6.258132	\\
2458689.76742	&	-4770.629672	&	6.004815	\\
2458690.79475	&	-4762.125532	&	4.466280	\\
2458691.77767	&	-4738.547657	&	4.140452	\\
2458750.58685	&	-4728.338615	&	4.929043	\\
2458752.61872	&	-4776.305472	&	4.542711	\\
2458753.60754	&	-4790.955086	&	4.069793	\\
2458754.65281	&	-4775.853945	&	4.689729	\\
2458755.65078	&	-4745.048769	&	3.621645	\\
2459163.62510	&	-4733.510345	&	4.777836	\\
2459178.57572	&	-4754.916775	&	4.606471	\\
2459181.59995	&	-4730.861963	&	4.374840	\\
2459182.59757	&	-4747.139606	&	4.972627	\\
2459184.60398	&	-4775.425414	&	4.197057	\\
\hline
\end{tabular}
\end{center}
\end{table}

\section{Analysis and Results}
\label{sec:3}
\subsection{Stellar Parameters}
\label{sec:3.1}
We collected stellar spectra on UT2018-11-16 with the 1.5-metre Tillinghast Refactor Echelle Spectrograph (TRES) \citep{2011ASPC..442..305M} at the Fred Lawrence Whipple Observatory (FLOW) in southern Arizona. The TRES reconnaissance spectra were organised by Dave Latham and collected by Gilbert Esquerdo. The spectra (SNR = 28.2) were extracted through a custom pipeline presented in \citet{buchhave2010}, and stellar parameters were determined using stellar parameter classification (SPC) tool \citep{buchhave2012,buchhave2014}. The stellar effective temperature $T_{eff}$ = 4994 $\pm$ 50 K, logarithmic surface gravity $logg_*$ = 4.685 $\pm$ 0.1, stellar rotational velocity $vsini$ = 2.28 $\pm$ 0.5 km $s^{-1}$, and metal abundance [m/H] of 0.272 $\pm$ 0.08 were obtained via this tool.

\begin{figure}
    \centering
    \includegraphics[scale = 0.3]{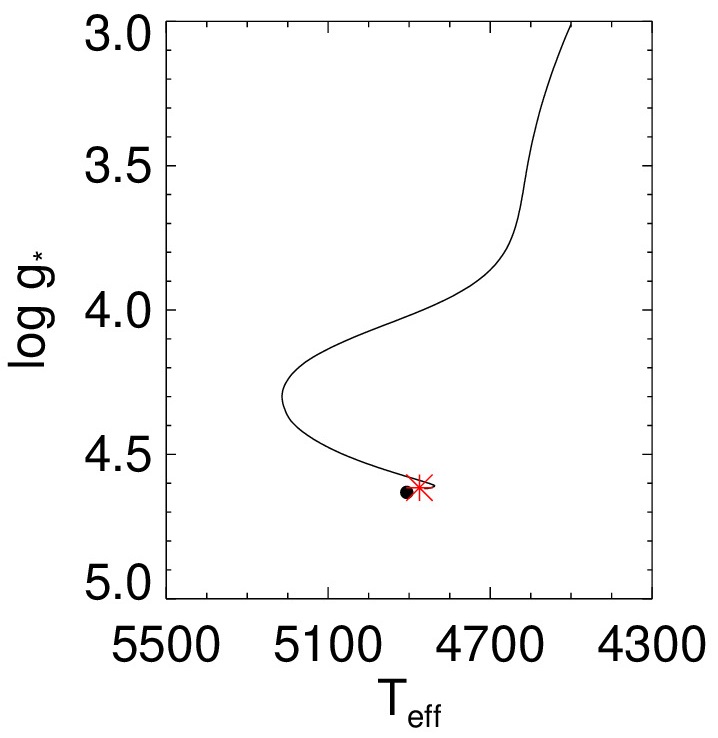}
    \caption{Stellar evolution track for TOI-181. Black line represents the most probable evolutionary track, black dot is a representation of model effective temperature and logarithmic surface gravity. Red asterisk symbol shows the equal evolutionary point or age of TOI-181.}
    \label{fig:lifespan}
\end{figure}

\begin{figure}
    \centering
    \includegraphics[scale = 0.25]{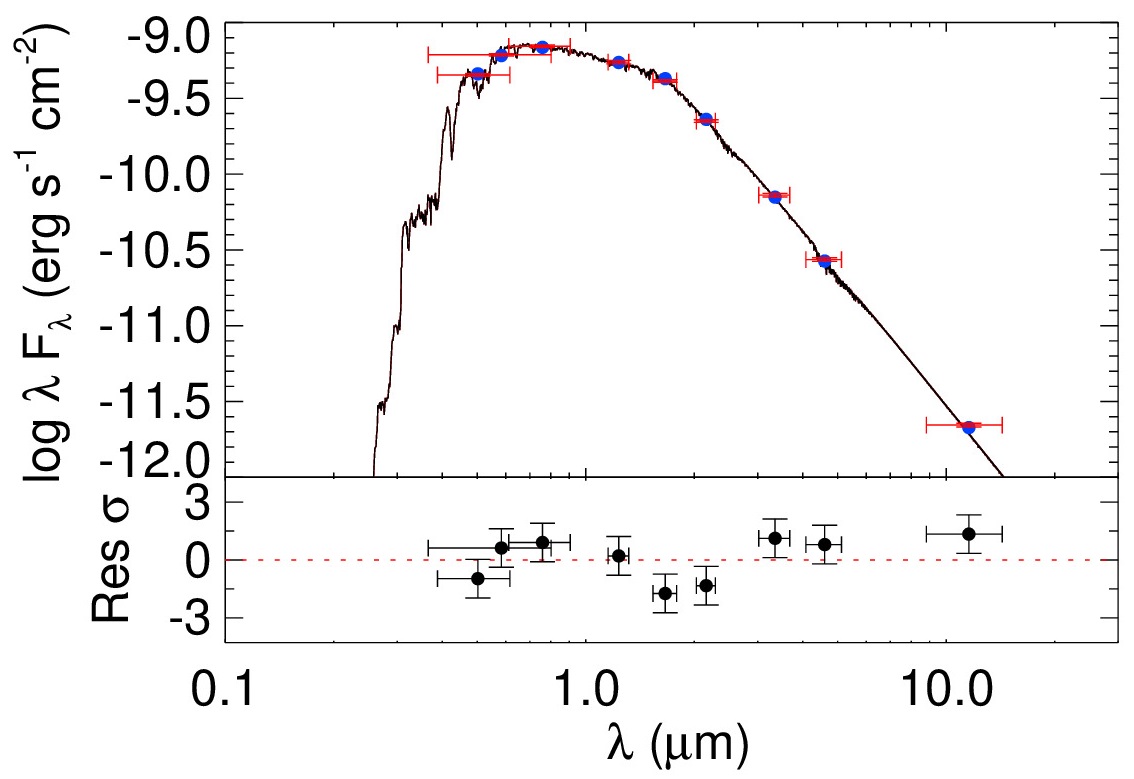}
    \caption{The spectral energy distribution of TOI-181. Red horizontal error bars show the broadband photometry measurements and vertical bars represent the uncertainty in measurements. Black curve shows the best fit Kurcuz stellar atmosphere model with blue circles representing model fluxes over each band.}
    \label{fig:sed}
\end{figure}

We derived stellar parameters by using spectral energy distribution (SED) \citep{2016ApJ...831L...6S} and MIST isochrones \citep{2016ApJ...823..102C,2016ApJS..222....8D} joint fits with EXOFASTv2 \citep{2019arXiv190709480E}. This joint fit can precisely measure the radius, mass, age, and $logg_*$ of the star \citep{2022arXiv220914301E}. To perform this framework, we used TESS transit photometry data, broadband photometry, and Gaia DR2 parallax \citep{2018A&A...616A...1G}. We placed priors on effective temperature and metallicity, determined using spectral analysis. We also showed the MIST stellar evolution track for TOI-181 in Fig \ref{fig:lifespan}. The best-fit SED model along with broadband photometry is shown in Fig. \ref{fig:sed} Through this analysis, we measured stellar parameters, $R_* = 0.745^{+0.024} _{-0.023} R_{\sun}$, $M_* = 0.822^{+0.040} _{-0.038}M_{\sun}$, Age = $5.4^{+4.9} _{-3.7}$ Gyr, and $logg_* = 4.608^{+0.035} _{-0.034}$. Table \ref{tab:stepa} shows the adopted parameters.

\begin{table}
\caption{Stellar Parameters}
\label{tab:stepa}
\begin{center}
\begin{tabular}{l c c}
\hline
Parameter & Unit & Value \\
\hline
\hline
RA & deg & 352.17${\degr}$ \\
DEC & deg & -34.49${\degr}$\\
Distance & pc & 96.23 $\pm$ 0.57 \\
Stellar Radius & $R_{\sun}$ & 0.745 $\pm$ 0.024 \\
Stellar Mass & $M_{\sun}$ & 0.822 $\pm$ 0.040 \\
Effective Temperature & K & 4994 $\pm$ 50 \\
Stellar Density & g $cm^{-3}$ & 2.80 $\pm$ 0.31 \\
Luminosity & $L_{\sun}$ & 0.298 $\pm$ 0.011 \\
Gravity & $logg_*$ & 4.608 $\pm$ 0.035 \\
Stellar Rotation & km $s^{-1}$ & 2.28 $\pm$ 0.50 \\ 
Metallicity & [m/H] & 0.272 $\pm$ 0.08 \\
Age & Gyr & 5.4 $\pm$ 4.9 \\
\hline
\multicolumn{3}{c}{Magnitudes}\\
 & & \\
TESS&	10.46	$\pm$	0.006	\\
B	&	12.23 	$\pm$	0.211	\\
V	&	11.19 	$\pm$	0.019	\\
Gaia&	11.08   $\pm$	0.001	\\
J	&	9.62 	$\pm$	0.029	\\
H	&	9.14 	$\pm$	0.022	\\
K	&	9.05 	$\pm$	0.021	\\
WISE 3.4 $\mu m$	&	8.96 	$\pm$	0.02	\\
WISE 4.6$\mu m$	&	9.04 	$\pm$	0.02	\\
WISE 12$\mu m$	&	8.93 	$\pm$	0.03	\\
WISE 22	$\mu m$&	8.48 	\\
\hline
\multicolumn{3}{c}{Other Identifiers}\\
 & & \\
\multicolumn{3}{c}{TIC 76923707}\\
\multicolumn{3}{c}{2MASS J23284126-3429289}\\
\multicolumn{3}{c}{APASS 17429793}\\
\multicolumn{3}{c}{Gaia DR2 6552346035980010624}\\
\multicolumn{3}{c}{TYC 7517-00248-1}\\
\multicolumn{3}{c}{UCAC4 278-232062}\\
\multicolumn{3}{c}{WISE J232841.35-342929.0}\\
\hline
\end{tabular}
\end{center}
\end{table}

Because the proper motion for TOI-181 is relatively large ($\mu_\alpha=111.3563$, $\mu_\delta=-2.6423$ mas/yr), we have checked archival images to search for a possible line-of-sight star that may currently be behind the star.  Utilizing the Second Palomar Sky Survey \citep{1991PASP..103..661R} and the 2MASS Survey \citep{2006AJ....131.1163S} and the Finder Chart service at IRSA linked through the ExoFOP \footnote{\url{https://irsa.ipac.caltech.edu/applications/finderchart/servlet/api?mode=getResult&locstr=352.17254091369102+-34.491409902166303&subsetsize=5&survey=DSS\%2CSDSS\%2C2MASS\%2CWISE}}, we searched for possible background contamination.  The oldest image is from the UK Schmidt survey in 1981, which is 40 years before the TESS observations.  In that time, TOI-181 has moved approximately 4\arcsec\ to the East (and 0.1\arcsec\ to the South).  The resolution of the 1981 imaging is approximately 5\arcsec, which would be sensitive to another approximately 2-3 magnitudes fainter and would appear as a shoulder on the point spread function of TOI-181.  There is no evidence in the imaging for the presence of another star at the current location of TOI-181.

Further, in support of TOI-181 being a single star, the Gaia solution is consistent with the star being single. Gaia DR3 astrometry \citep{2021A&A...650C...3G} provides additional information on the possibility of inner companions that may have gone undetected by either Gaia or the high resolution imaging.  The Gaia Renormalised Unit Weight Error (RUWE) is a metric, similar to a reduced chi-square, where values that are $\lesssim 1.4$  indicate that the Gaia astrometric solution is consistent with the star being single whereas RUWE values $\gtrsim 1.4$ may indicate an astrometric excess noise, possibly caused the presence of an unseen companion \citep[e.g., ][]{2020AJ....159...19Z}.  TOI-181 has a Gaia DR3 RUWE value of 0.88 indicating that the astrometric fits are consistent with the single star model.

\subsection{Periodogram Analysis}
\label{sec:3.2}
\begin{figure*}
    \centering
    \includegraphics[scale = 0.45]{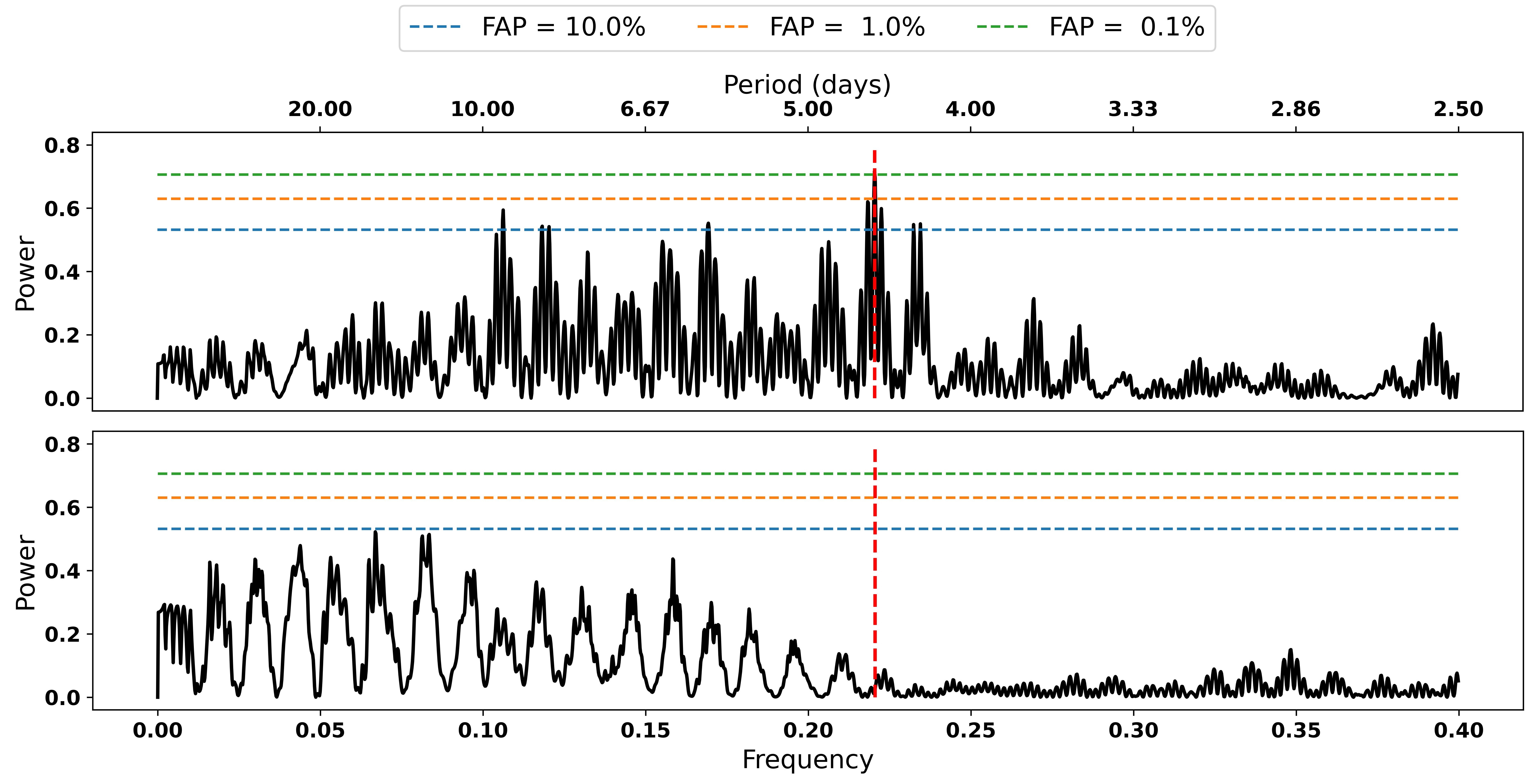}
    \caption{The GLS periodogram for RVs (upper panel) and residuals RVs (lower panel). The primary peak is observed at $\approx$ 4.53 days (red dashed line). This period is consistent with orbital period derived using transit photometry. FAP levels of 0.1 \%, 1 \% and 10 \% for the periodogram are shown in legend.}
    \label{fig:rvperiod}
\end{figure*}

We generated a Generalized Lomb-Scargle (GLS) periodogram \citep{2009A&A...496..577Z} for RVs and residual RVs. Fig. \ref{fig:rvperiod} shows the resulting periodogram. We started by subtracting the corresponding mean RV from RV values to remove offset. The periodogram is plotted between the lowest frequency 0.0 $day^{-1}$ and highest frequency measured using Nyquist limit for unevenly spaced data \citep{2018ApJS..236...16V}, which resulted in $\approx$ 0.40 $day^{-1}$. We calculated the false alarm probability (FAP) using the equation presented in the \citet{2009A&A...496..577Z}. Any significant signal is considered to have a FAP threshold of 0.1\% or smaller. The top panel of Fig. \ref{fig:rvperiod} shows that a significant peak is found at $\approx$ 4.53 days (red dashed line indication) with a FAP of 0.086 \%. This period is consistent with the measured period using transit photometry. Bottom panel of Fig. \ref{fig:rvperiod} shows the periodogram of residual RVs, which has no significant peak above the considered threshold, which indicates the absence of any other periodic signal in the data, noting the observation capabilities of the instrument.

\subsection{Validation using FPP}
\label{sec:3.3}
We employed two statistical validation tools, \texttt{VESPA} \citep{2012ApJ...761....6M} and \texttt{TRICERATOPS} \citep{2020ascl.soft02004G} as a preliminary check for the exoplanet. We make use of \texttt{VESPA} version 0.6 through a virtual machine created by Kevin Hardegree-Ullman\footnote{\url{https://github.com/kevinkhu/vespa-vm}}. To calculate FPP for any given target, \texttt{VESPA} requires three input files, \texttt{star.ini} that contain the parameters (effective temperature, metallicity, logg, magnitude of the star in different bands, including the TEES band) of the target star. \texttt{fpp.ini} file contains the orbital period, planet-to-star radius ratio, maximum aperture radius (in arcsec), maximum permitted secondary depth of a potential secondary eclipse (\texttt{secthresh}), position of target (RA and DEC), telescope filter, and cadence (in days). \texttt{transit.txt} file that contains the data from transit photometry. If the calculated FPP for the target is $<$ 1 \% \citep{2012ApJ...761....6M,2016ApJ...822...86M}, then it can be validated as a planet. In Table \ref{tab:vesparesult}, we have compiled a list of all \texttt{VESPA} outputs.

\begin{table}
\caption{\texttt{VESPA} output for all FP scenarios. Shows the calculated probability of various scenarios.}
\label{tab:vesparesult}
\begin{center}
\begin{tabular}{l c c c}
\hline
Scenario & Sector 02 & Sector 29 & Combined \\
\hline
\hline
$\mathcal{P}$(HEB)	&	9.04$\times 10^{-10}$	&	6.49$\times 10^{-8}$	&	1.35$\times 10^{-17}$	\\
$\mathcal{P}$(HEB) ($\times$2P)	&	2.92$\times 10^{-14}$	&	1.38$\times 10^{-11}$	&	1.67$\times 10^{-14}$\\
$\mathcal{P}$(EB)	&	5.56$\times 10^{-13}$ &	8.05$\times 10^{-20}$	&	2.83$\times 10^{-20}$	\\
$\mathcal{P}$(EB) ($\times$2P)	&	7.03$\times 10^{-33}$	&	1.79$\times 10^{-25}$	&	6.78$\times 10^{-33}$	\\
$\mathcal{P}$(BEB)	&	9.67$\times 10^{-20}$ &	5.66$\times 10^{-32}$	&	5.66$\times 10^{-17}$ \\
$\mathcal{P}$(BEB) ($\times$2P)	&	0.00	&	0.00	&	0.00	\\
$\mathcal{P}$(Planet)	&	1.00	&	1.00	&	1.00	\\
\hline
FPP	&	9.05$\times 10^{-10}$	&	6.50$\times 10^{-8}$	&	1.68$\times 10^{-14}$	\\
\hline
\end{tabular}
\end{center}
\end{table}

The results of the calculations for FPP from \texttt{VESPA} are significantly lower than the threshold of 1 \%. It presents a convincing illustration of a target being considered a planet.

The TOI-181 signal was put through a different statistical validation software called \texttt{TRICERATOPS} \citep{2020ascl.soft02004G} so that we could conduct an independent check. Employing an algorithm  similar  to \texttt{VESPA}, It investigates a variety of astrophysical situations that could produce false-positive signals. \texttt{TRICERATOPS} was developed in particular for the purpose of validating TESS observations. The aperture mask of a target, the transit depth, and the orbital period are all required as priors. Based on the parameters provided, it uses a Bayesian framework to check for the various FP scenarios. The added advantage over \texttt{VESPA} is that it also performs an analysis not only on the target star but also on the stars that are located within a 2.5' radius. And using this information, it computes both the FPP of the signal as well as the probability that the planet candidate is a false positive originating from a known nearby star. This probability is referred to as the nearby FPP (NFPP).

\texttt{TRICERATOPS} is able to confirm that a signal is a planet if the FPP < 0.015 and the NFPP < 0.001, as stated in \citet{2021AJ....161...24G}. If the FPP < 0.5 and the NFPP is < 0.001, it is probably a planet; however, if the NFPP is greater than 0.1, the signal is probably due to a nearby false positive scenario. Table \ref{tab:fpptri} contains a compilation of the FPP that was calculated for TOI 181 using \texttt{TRICERATOPS} for each of the different sectors.

\begin{table}
\caption{FPP and NFPP calculated using \texttt{TRICERATOPS}}
\label{tab:fpptri}
\begin{center}
\begin{tabular}{l c c}
\hline
Sector & FPP & NFPP \\
\hline
\hline
02 & 0.0006061 $\pm$ 0.0027549 & 4.2022$\times 10^{-28}$ $\pm$ 6.1927$\times 10^{-28}$\\
29 & 0.0022743 $\pm$ 0.0053222 & 4.0302$\times 10^{-31}$ $\pm$ 8.7789$\times 10^{-28}$\\
Both & 0.0003812 $\pm$ 0.0009869 & 2.5397$\times 10^{-53}$ $\pm$  9.1340$\times 10^{-53}$\\
\hline
\end{tabular}
\end{center}
\end{table}

\subsection{Planetary Parameters}
\label{sec:3.4}
We employed the \texttt{Juliet} \citep{2019MNRAS.490.2262E} Python package to model the transit and radial velocity data. The \texttt{Juliet} is a versatile modelling tool for extra-solar planetary systems that allows the fast and easy calculation of parameters from transit photometry, radial velocity, or both using Bayesian inference and, in particular, Nested Sampling to allow both effective measurement and suitable model comparisons. \texttt{Juliet} can take various datasets of transit photometry and radial velocity simultaneously and calculate systematic trends using linear models or Gaussian Processes (GP). The priors that we used during the modelling process are displayed in Table \ref{tab:priors}.

\begin{table}
\caption{Priors provided to \texttt{Juliet} for modeling}
$\mathcal{N}$: Normal Distribution\\
$\mathcal{U}$: Uniform Distribution\\
$\mathcal{L}$: Log-uniform Distribution\\
\label{tab:priors}
\begin{center}
\begin{tabular}{l c c}
\hline
Prior & Description & Distribution \\
\hline
\hline
Period (P) & days & $\mathcal{N}$(4.53193, 0.1) \\
$T_0$ & BJD & $\mathcal{N}$(2458358.1182, 0.1) \\
$r_1$ & \citet{2018RNAAS...2..209E} & $\mathcal{U}$(0.0, 1.0)\\
$r_2$ & \citet{2018RNAAS...2..209E} & $\mathcal{U}$(0.0, 1.0)\\
a/$R_*$ & & $\mathcal{U}$(10, 20)\\
K & m s$^{-1}$ & $\mathcal{U}$(15, 23)\\
Eccentricity & & $\mathcal{U}$(0.0, 0.7)\\
$\omega$ & deg & $\mathcal{U}$(-360, 360)\\
 & & \\
\multicolumn{3}{c}{Instrumental Parameters}\\
\hline
 & & \\
$q_1$ & \citet{2013MNRAS.435.2152K} & $\mathcal{U}$(0.0, 1.0)\\
$q_2$ & \citet{2013MNRAS.435.2152K} & $\mathcal{U}$(0.0, 1.0)\\
$m_{flux}$ & ppm & $\mathcal{N}$(0.0, 0.1)\\
$m_{dilution}$ & & 1.0\\
$\sigma_{w}$ & ppm & $\mathcal{L}$(0.1, 1000)\\
$\mu_{instrument}$ & m $s^{-1}$ & $\mathcal{U}$(a, b)\\
\hline
\end{tabular}
\end{center}
\end{table}

\begin{figure}
\begin{center}
\includegraphics[scale=0.45]{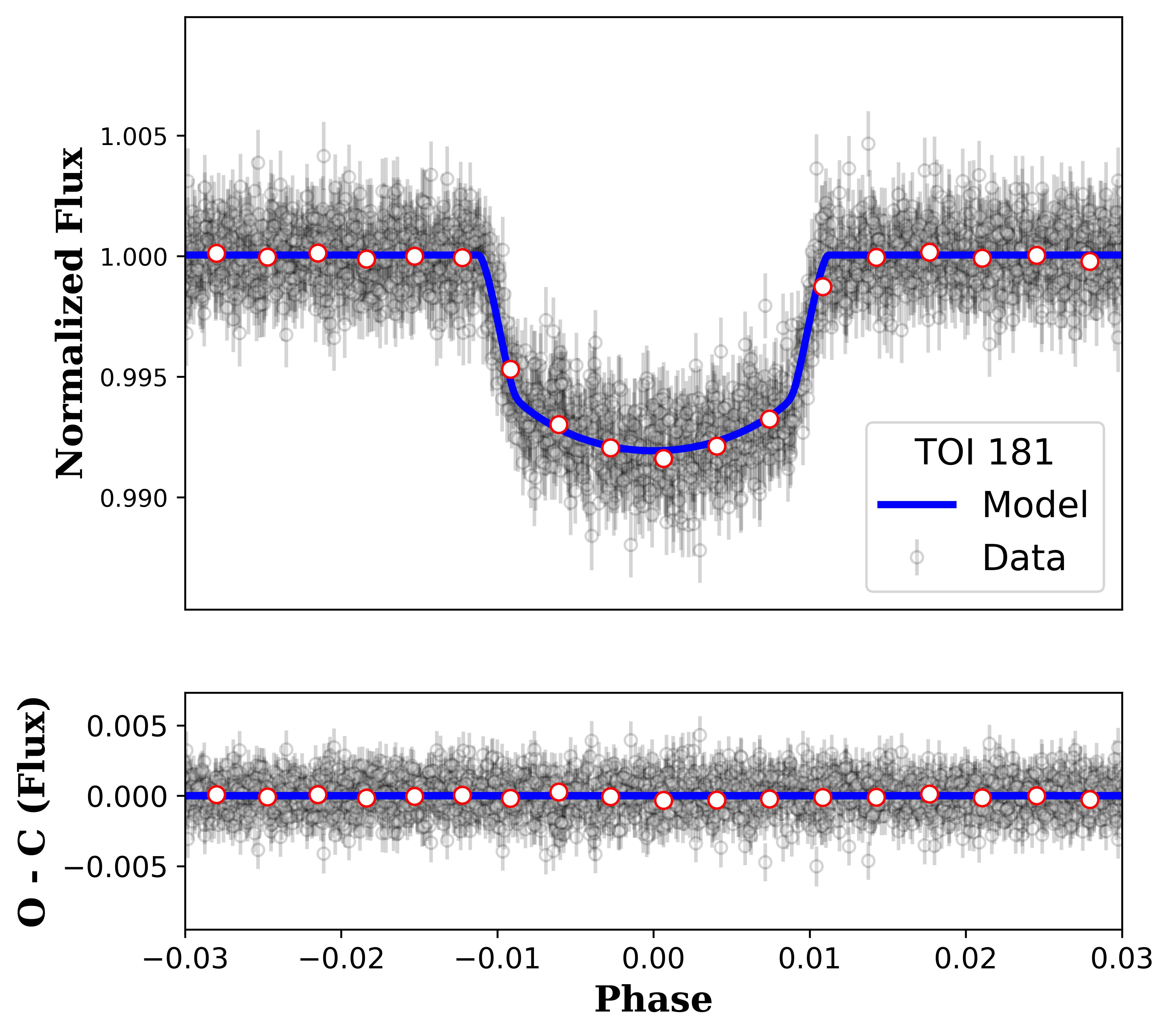}
\end{center}
\caption{Phase folded transit light curve centred on T0 (BTJD 2458358.1180). Blue line suggests the best-fit transit model with log evidence ($lnz$) of 171216.32083 $\pm$ 0.44260 using \texttt{Juliet}. Bottom panel shows the residuals between the best-fit model and data. Red white-filled markers shows the binned data.}
\label{fig:transitphase}
\end{figure}

\begin{figure}
\begin{center}
\includegraphics[scale=0.45]{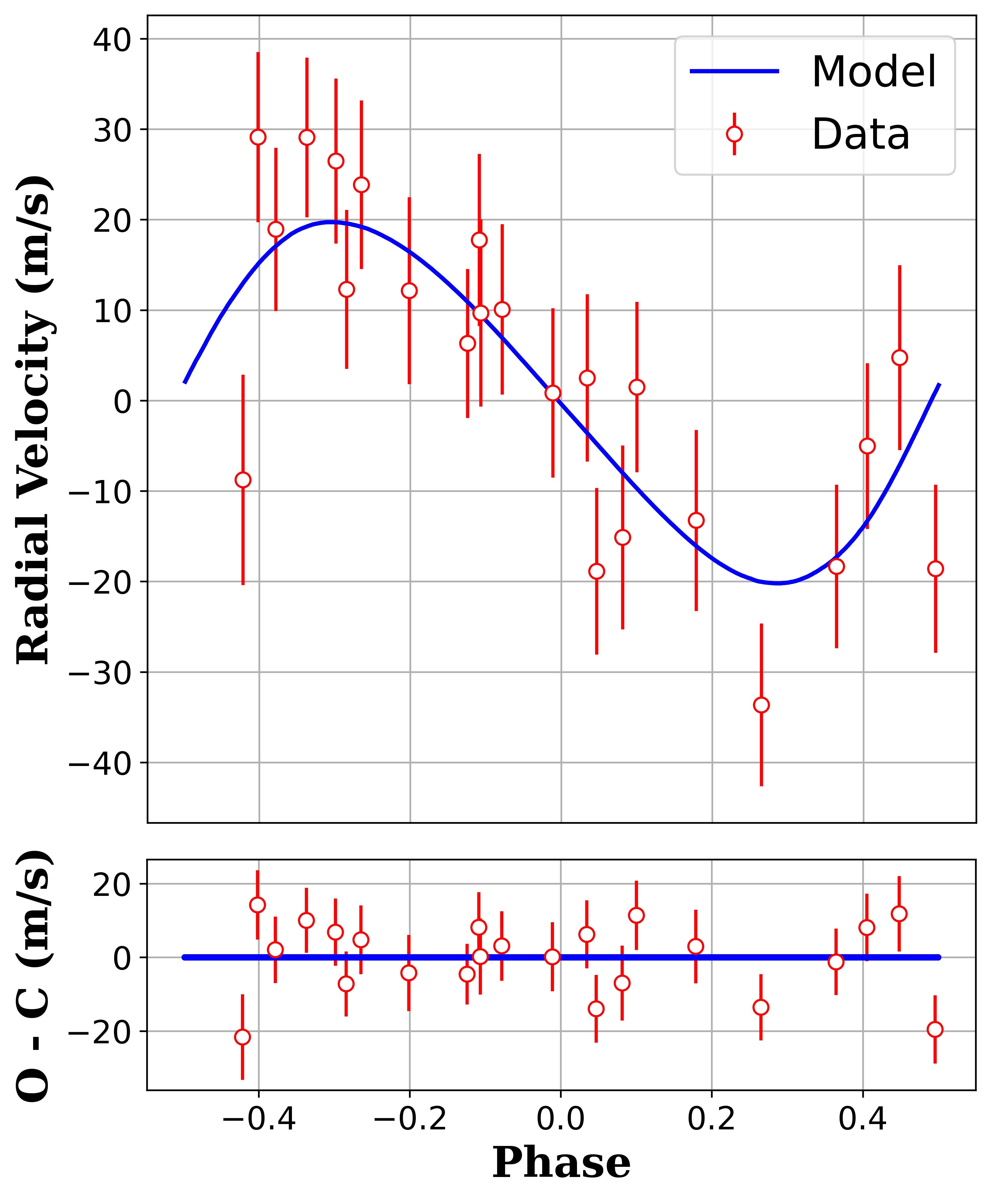}
\end{center}
\caption{Phase folded radial velocity curve, based on data obtained from the ESO Science Archive Facility with DOI: \url{https://doi.org/10.18727/archive/33}. Red line suggests the best-fit radial velocity model using \texttt{Juliet}. Bottom panel shows the residuals between the best-fit model and data.}
\label{fig:rvtphase}
\end{figure}

From the best-fit model of transit (Fig. \ref{fig:transitphase}) and radial velocity (Fig. \ref{fig:rvtphase}) using \texttt{Juliet}, we calculated planetary and orbital parameters. Based on this TOI-181b is a planet having the density 0.75 g cm$^{-3}$, orbiting its host star in a slightly eccentric orbit of eccentricity 0.15 having semi-major axis of 0.054 AU with orbital period 4.5320 days. TOI-181b has radius of 0.6339 R$_J$ and mass of 0.1452 M$_J$ and has equilibrium temperature (assuming Bond Albedo (A$_B$) = 0) 895.10 K. As mentioned mass of TOI-181b is 0.1452 M$_J$ which is well-below the exoplanet mass-limit of 13 M$_J$ \citep{2022NewAR..9401641L}, so along with measured FPPs using \texttt{VESPA} and \texttt{TRICERATOPS}, radial velocity measurements also rules out the possibility of eclipsing binary and confirms the planetary nature of TOI-181b. Using the \texttt{Juliet} best-fit model we also derived quadratic limb darkening parameters for TESS infrared band: $u_1$ = 0.7259 $^{+ 0.08166 } _{- 0.1187 }$, $u_2$ = -0.28075 $^{+ 0.1749 } _{- 0.0870 }$.

Based on the priors provided, all the calculated planetary and orbital parameters (mean value and 68 \% confidence interval of the posterior probability distribution) are listed in Table \ref{tab:finalpara}.

\renewcommand{\arraystretch}{1.5}
\begin{table}
\begin{center}
\caption{Median values and 68\% confidence interval for TOI 181 b from \texttt{Juliet}}
\label{tab:finalpara}
\begin{tabular}{l l c}
\hline
\hline
Parameters & Description (Unit) & Values\\
\hline
 P\dotfill & Period (days)\dotfill & 4.5320 $^{+ 0.000002 } _{- 0.000002 }$ \\
 $R_P$\dotfill & Radius ($R_J$)\dotfill & 0.6339 $^{+ 0.0079 } _{- 0.0092 }$ \\
 $R_P$\dotfill & Radius ($R_{\earth}$)\dotfill & 6.9559 $^{+ 0.087 } _{- 0.101 }$ \\
 $T_C$\dotfill & Epoch Time (BJD)\dotfill & 2458358.12 $^{+ 0.00027 } _{- 0.00026 }$ \\
 $T_d$\dotfill & Transit Duration (days)\dotfill & 0.0829 $^{+ 0.0097 } _{- 0.0088 }$ \\
 $a$\dotfill & Semi-major Axis (AU)\dotfill & 0.0539 $^{+ 0.0043 } _{- 0.0036 }$ \\
 $i$\dotfill & Inclination (Degree)\dotfill & 88.28 $^{+ 0.32 } _{- 0.34 }$ \\
 $e$\dotfill & Eccentricity\dotfill & 0.1543 $^{+ 0.06 } _{- 0.03 }$ \\
 $\omega$\dotfill & Argument of Periastron (Degree)\dotfill & -96.91 $^{+ 19.48 } _{- 28.82 }$ \\
 $T_{eqq}$\dotfill & Equilibrium Temperature (K)\dotfill & 895.10 $^{+ 32.14 } _{- 33.86 }$ \\
 $S$\dotfill & Insolation ($S_{\earth}$)\dotfill & 102.43 $^{+ 15.52 } _{- 14.64 }$ \\
 $M_P$\dotfill & Mass ($M_J$)\dotfill & 0.1452 $^{+ 0.0085 } _{- 0.024 }$ \\
 $M_P$\dotfill & Mass ($M_{\earth}$)\dotfill & 46.1687 $^{+ 2.71 } _{- 7.83 }$ \\
 K\dotfill & RV Semi-amplitude (m $s^{-1}$)\dotfill & 20.5610 $^{+ 1.46 } _{- 2.37 }$ \\
 $logK$\dotfill & Log of RV Semi-amplitude\dotfill & 1.3130 $^{+ 0.029 } _{- 0.053 }$ \\
 $R_P/R_S$\dotfill & Radius of planet in stellar radii\dotfill & 0.0854 $^{+ 0.001 } _{- 0.001 }$ \\
 $a/R_S$\dotfill & Semi-major axis in stellar radii\dotfill & 15.5641 $^{+ 1.25} _{- 1.06 }$ \\
 $\delta$\dotfill & Transit Depth (Fraction)\dotfill & 0.0073 $^{+ 0.00018 } _{- 0.00021 }$ \\
 $b$\dotfill & Impact Parameter\dotfill & 0.5348 $^{+ 0.05 } _{- 0.07 }$ \\
 $\rho$\dotfill & Density (cgs)\dotfill & 0.7562 $^{+ 0.015 } _{- 0.100 }$ \\
 $g_P$\dotfill & Surface Gravity (m $s^{-2}$)\dotfill & 9.3642 $^{+ 0.30 } _{- 1.35 }$ \\
 $M_P \sin i$\dotfill & Minimum Mass ($M_J$)\dotfill & 0.1452 $^{+ 0.008 } _{- 0.024 }$ \\
 $M_P/M_S$\dotfill & Mass Ratio\dotfill & 0.000169 $^{+ 0.00001 } _{- 0.00003 }$ \\
 \hline
 \hline
 \multicolumn{3}{c}{TESS Infrared Band (Quadratic Limb Darkening Parameters)}  \\
 $u_1$\dotfill & Limb Darkening Parameter\dotfill & 0.6888 $^{+ 0.10 } _{- 0.13 }$ \\
 $u_2$\dotfill & Limb Darkening Parameter\dotfill & -0.2350 $^{+ 0.21 } _{- 0.12 }$ \\
 \hline
\end{tabular}
\end{center}
\end{table}

\section{Discussion}
\label{sec:4}
This paper describes the discovery of TOI-181b, the first exoplanet confirmed by the VaTEST project. To further our understanding of sub-Saturn planets orbiting metal-rich stars, precise determination of host star and planet properties of TOI-181b is invaluable. In our discussion, we focus on derived planet properties and their prominent features that make the planet interesting to study.

\subsection{TOI-181b and other known sub-Saturns}
 TOI-181b has a radius of 0.6339 R$_J$ and a mass of 0.1452 M$_J$, which are greater than Neptune's radius and mass but smaller than Saturn's, placing it in the sub-Saturn or super-Neptune group. The density measured from mass and radius is 0.75 g cm$-3$, which is comparable with Saturn's density (0.69 g cm$^{-3}$). \citet{2017AJ....153..142P} has gathered a sample of 23 sub-Saturns (Table \ref{tab:A1}) with densities measured at 50\% or better. Further this sample was extended by 7 more sub-Saturns (Table \ref{tab:A2}) by \citet{2020MNRAS.497.4423N}. We expanded this sample with the discovery of TOI-181b and also identified 16 more sub-Saturns (Table \ref{tab:A3}) with measured densities with a precision of 50\% or better. K2-98b \citep{2016AJ....152..193B}, planet having radius 4.3 $R_{\earth}$ and mass 32 $M_{\earth}$, which is one of the dense sub-Saturns ($\rho$ = 2.15 g $cm^{-3}$) after Kepler-413b \citep[$\rho$ = 3.2 g $cm^{-3}$]{2014ApJ...784...14K} and K2-108b \citep[$\rho$ = 2.22 g $cm^{-3}$]{2017AJ....153..142P}. TOI-421 c \citep{2020AJ....160..114C} having radius 5.09 $R_{\earth}$ and mass 16.42 $M_{\earth}$, which is one of the few sub-Saturns orbiting in multi-planetary system. An ultra-hot sub-Saturn LTT 9779 b \citep{2020NatAs...4.1148J} having equilibrium temperature 1978 K. Other sub-Saturns are listed in Table \ref{tab:A3}.
 
\begin{figure}
\begin{center}
\includegraphics[scale=0.43]{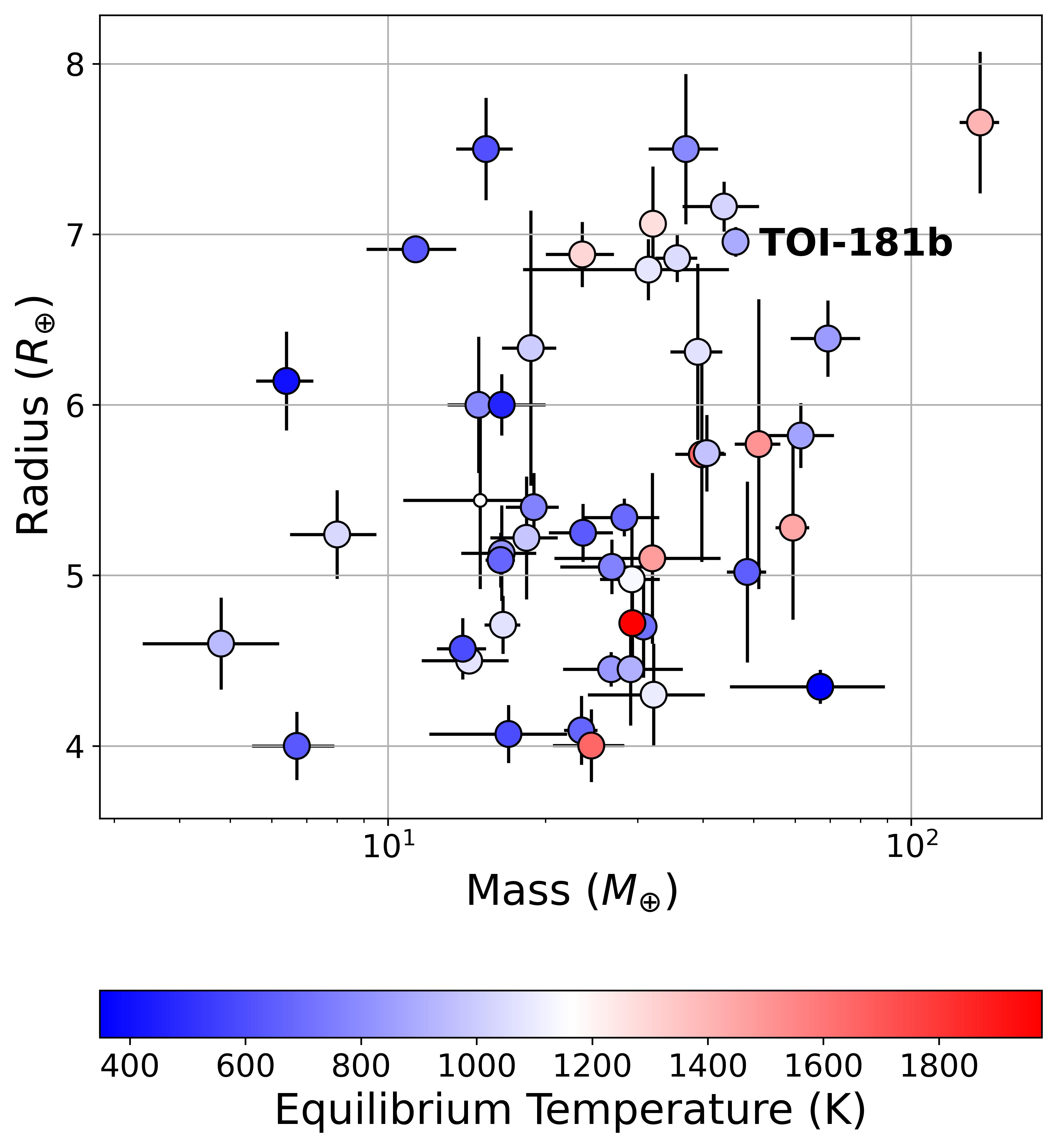}
\includegraphics[scale=0.43]{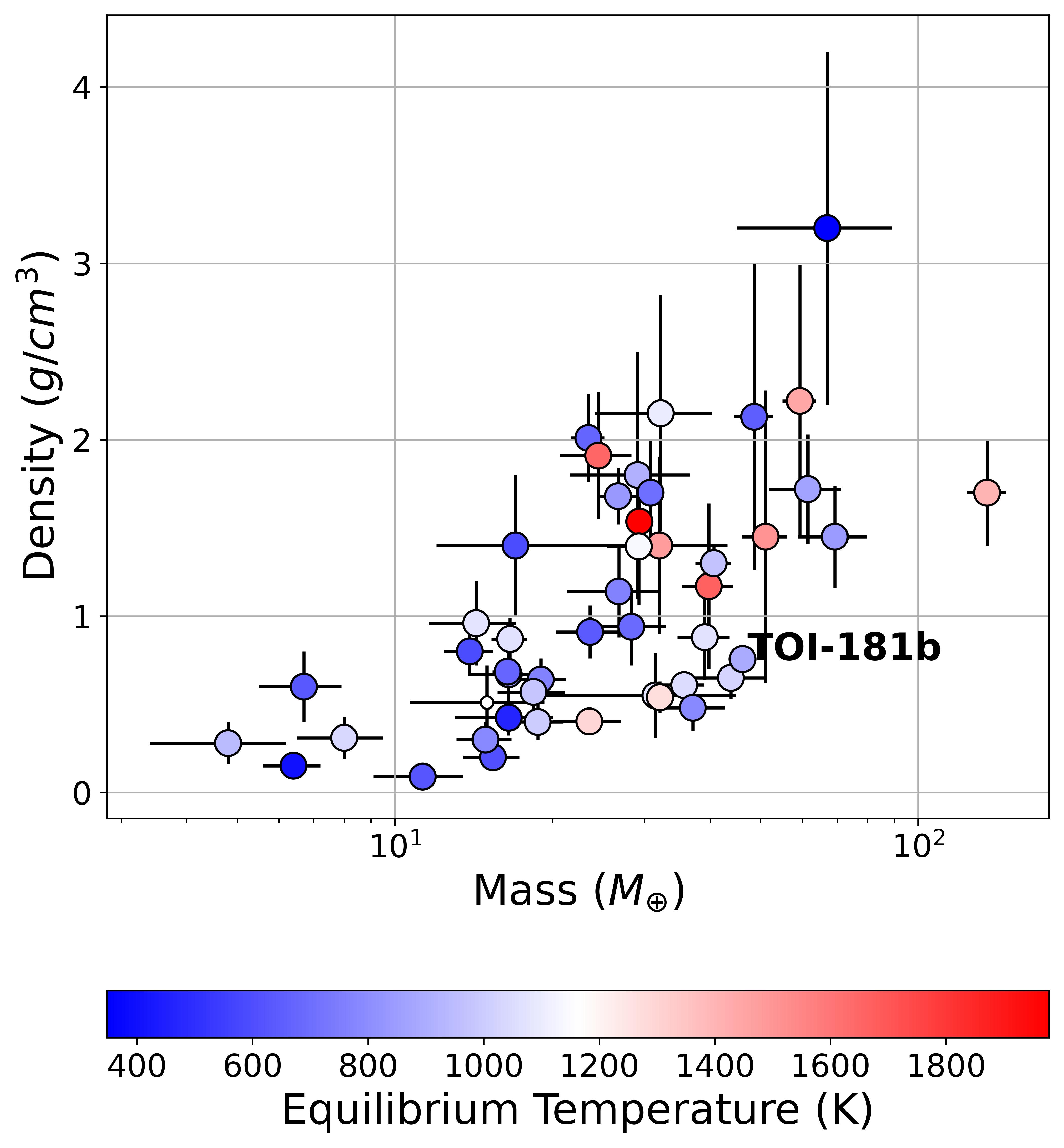}
\end{center}
\caption{Top figure: Scatter plot of planetary radius and mass of all the known sub-Saturns. Bottom figure: Scatter plot of planetary density (g cm$^{-3}$) vs planetary mass ($M_{\earth}$). Colorbar shows the equilibrium planetary temperature in kelvin. TOI-181b is marked by text. Planets are taken from the NASA Exoplanet Archive (\url{https://exoplanetarchive.ipac.caltech.edu/}).}
\label{fig:dvsm}
\end{figure}
 
The top panel of Fig. \ref{fig:dvsm} shows the scatter plot for radius and mass of all known sub-Saturns along with TOI-181b. This plot depicts the main characteristics of sub-Saturns observed by \citet{2017AJ....153..142P}. There seems to be no correlation between planetary radius and mass. This type of planets have a narrow range of radii (4–8 $R_{\earth}$) but is uniformly distributed throughout its mass. As can be seen from the Fig. \ref{fig:dvsm}, there is an absence of sub-Saturns for radii range 5.5 - 6.5 $R_{\earth}$ and masses of 20 - 30 $M_{\earth}$. TOI-181b is located at the upper edge of this scarcity, which makes it an interesting object to study the transition between small and massive sub-Saturns.

The bottom panel of Fig. \ref{fig:dvsm} shows the plot of density vs mass for the known sample of sub-Saturns. There seems to be little correlation between these parameters. For this 47 sub-Saturn sample, the Spearman's rank correlation coefficient is r = 0.62, which is a little lower than the r = 0.79 measured by \citet{2017AJ....153..142P} and the r = 72 measured by \citet{2020MNRAS.497.4423N}. This shows that the association between these two variables would be considered statistically significant.

\subsection{Eccentricity and Orbital Circularization}
TOI-181b has non-zero eccentricity of 0.15 and a short orbital period of 4.52 days. Based on this it can be expected that tides will damp away the eccentricity. To check this scenario we calculated orbit circularization rate \citep{1966Icar....5..375G},
\begin{equation}
\label{eq:1}
    \tau_e = \frac{4}{63}\left(\frac{Q'}{n}\right)\left(\frac{M_P}{M_*}\right)\left(\frac{a}{R_P}\right)^5
\end{equation}
where, n is mean motion given by $\sqrt{GM_*/a^3}$ and Q' is modified tidal quality factor which is given by $Q' = 3Q/2k_2$. Q is specific dissipation function and $k_2$ is Love number \citep{1966Icar....5..375G,2004ApJ...614..955M}. Value for $Q'$ is uncertain even for solar system planets. \citet{2017AJ....153..142P} had assumed $Q' = 10^5$ for sub-Saturns. We are following the same assumption here to measure circularization rate. As per eq. \ref{eq:1} we measured $\tau_e$ = 0.42 Gyr. Because this timescale is shorter than the age of its host star, the circularization caused by tides cannot account for the eccentric nature of planetary orbits. Though if we use $Q' = 10^6$ resultant $\tau_e$ is 4.19 Gyr which is greater than the lower bound of stellar age. So this tension can be alleviated by adopting $Q' = 10^6$ instead of $10^5$. Alternative solutions include additional undetected exoplanets in the system. This eccentric orbit can persist by the planet-planet interactions. So if this is true than it is very plausible that a second planet that is yet to be discovered is causing continual gravitational perturbations to keep TOI-181b's eccentricity constant over very long periods.

\subsection{Atmospheric Characterization}
By taking into account the brightness of the host star, radius, mass and planetary equilibrium temperature, \cite{2018PASP..130k4401K} developed a method to calculate Transmission and Emission Spectroscopy Metrics (TSM and ESM). TSM and ESM will make it simple to identify the best atmospheric characterization targets among the TESS planet candidates by calculating the expected James Webb Space Telescope (JWST) signal-to-noise ratio in transmission and thermal emission spectroscopy for a given planet. For TOI-181b TSM is measured as 179.17 which is above the threshold of forth quartile suggested by \cite{2018PASP..130k4401K}, that makes it a good target for transmission spectroscopy. ESM is measured as 34.76 which is also well above the threshold of 7.5 that makes the planet amenable to perform atmospheric characterization, with a focus on JWST capabilities. By assuming H/He atmosphere we calculated atmospheric scale height ($H = K_b T_{eq}/\mu g$ \citet{2014prpl.conf..739M}, where $\mu$ (mean molecular mass) = 2.4, $K_b$ (Boltzmann Constant)) of 331.13 km. Based on this value we estimated amplitude of spectral features for transmission spectroscopy \citep{2018haex.bookE.100K} which arrived to be 0.023\%. This suggest that TOI-181b may not suitable for ground-based transmission spectroscopy but the previous results of TSM and ESM suggesting it to be the good target for JWST spectroscopy. By assessing the elemental compositions and total metal enrichment of planet's atmosphere, such studies could directly evaluate our hypotheses regarding the planet's bulk composition and creation history. Future investigations will determine the atmospheric composition of this and other sub-Saturns with low densities, thereby shedding light on the nature of these mystery objects.

\subsection{Planetary Envelope Fraction}
\label{p_env}
Sub-Saturns are structurally distinct from other types of planets. It has both heavy elements in its core and a gaseous envelope surrounding the planet, both of which contribute significantly to the overall mass. Having such large radii imply that the envelopes of H/He make up a significant percentage of their overall mass \citep{2014ApJ...792....1L}, with typical envelope fractions ($f_{env}$) ranging from 10\% to 50\%. The \texttt{PLATYPOS} (PLAneTarY Photoevaporation Simulator) algorithm \citep{2022AN....34310105K} is used to calculate the envelope fraction and core mass of TOI-181b. This simulator estimates $f_{env}$ using the \citet{2014ApJ...792....1L} model, then uses existing parametrizations of planetary mass-radius relations with an energy-limited hydrodynamic escape model to estimate the mass-loss rate over time. We used the grid of core mass, radius, distance, and metallicity to solve for planetary mass. We interpolated the known planetary mass to estimate the $f_{env}$. Based on the assumption \citep{2016ApJ...818...36P} that the envelope fractions are unaffected by the detailed structure and composition of the core, we estimated the $f_{env}$ to be 60.97\%, implying that TOI-181b has the majority of its mass in its H/He envelope. We measured the core mass of 18.02 $M_{\earth}$ and core radius of 2.06 $R_{\earth}$ (super-Earth sized). Estimated $f_{env}$ for TOI-181b is highest among all known sub-Saturns. We calculated planetary metal enrichment ($Z_P/Z_*$ where, $Z_* = Z_{\sun}10^{[Fe/H]}$, $Z_P = M_{core}/M_{env}$) to be 17.28. This value is towards the lower end of the other planets' metal enrichment fraction \citep{2016ApJ...831...64T} as the maximum mass of TOI-181b is carried in its envelope. We calculated the planet's gravitational binding energy of $6.85 \times 10^{41}$ erg, for which the survival threshold predicted by photo-evaporation and thermal evolution models is $\approx 7.29 \times 10^{40}$ erg \citep{2012ApJ...761...59L}. For our target, we measured the lifetime integrated XUV flux incident to be $1.35 \times 10^{36}$ erg, which is well below the mentioned threshold. As a result, TOI-181b is either sufficiently massive or in a wide enough orbit to be immune to significant photo-evaporation.

\subsection{Hot-Neptune Desert}
\begin{figure}
\begin{center}
\includegraphics[scale=0.38]{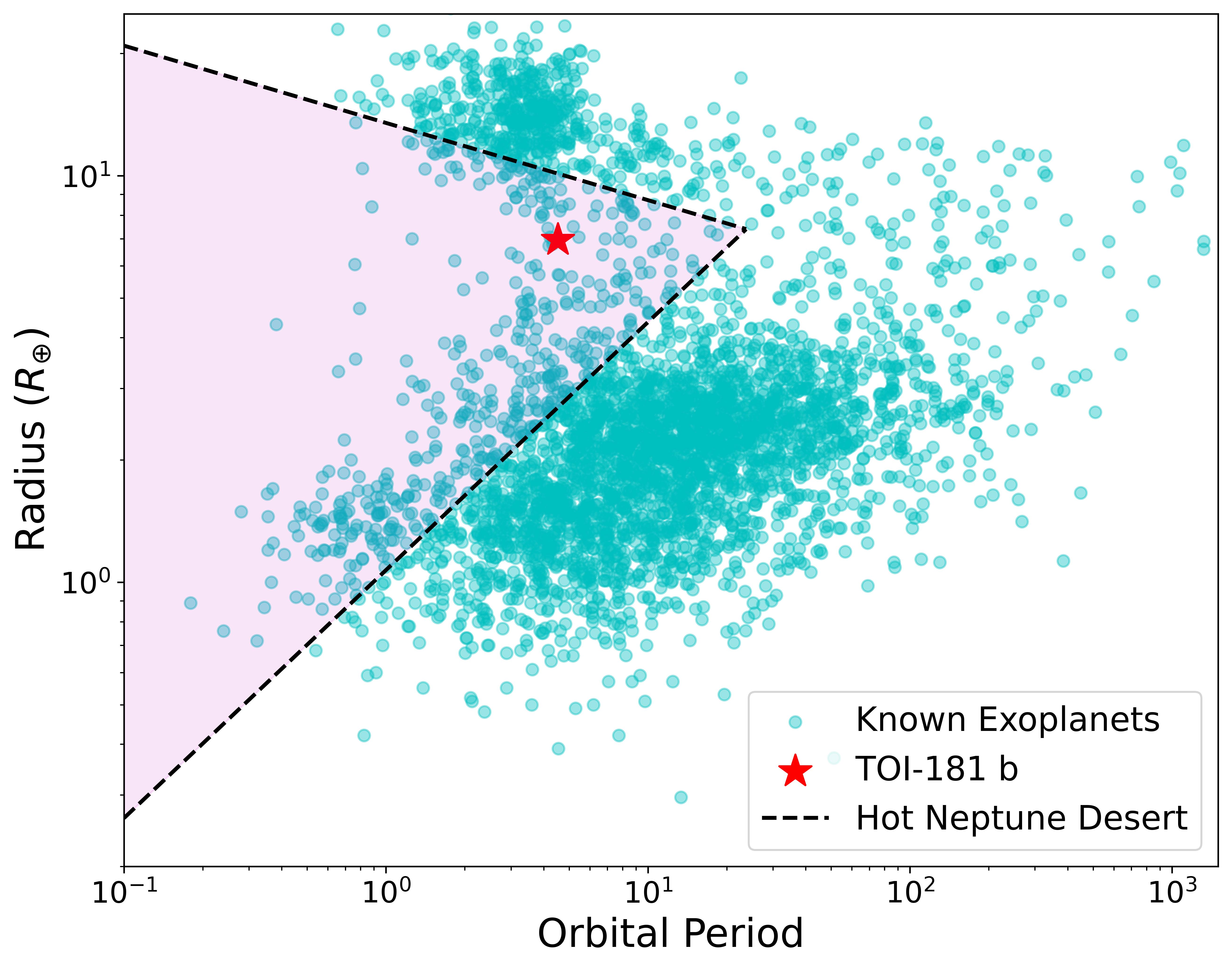}
\end{center}
\caption{Distribution of known exoplanets in the graph of radii vs orbital period. Red star sign shows TOI-181b and region enclosed inside black dashed line shows the "Hot Neptune Desert". Planets are taken from the NASA Exoplanet Archive (\url{https://exoplanetarchive.ipac.caltech.edu/})}
\label{fig:desert}
\end{figure}

Fig. \ref{fig:desert} depicts the relationship between planetary radii and orbital periods for all known exoplanets. TOI-181b carries a red star indication. The study demonstrates that Neptune-like planets in tight orbits are uncommon due to X-ray or Extreme Ultra Violet (EUV) emission from the host rapidly stripping close-in short-period planets' atmospheres and leaving behind low-mass rocky cores \citep{2013ApJ...776....2L,2018MNRAS.479.5012O,2016ApJ...831..180C}. This region is known as the "Hot Neptune Desert" \citep{2016A&A...589A..75M}. The dashed line in Fig. \ref{fig:desert}  depicts the Neptune desert. The suggested photoevaporation mechanism is consistent with the lowest boundary of the hot Neptune desert (i.e., planets with small radii). In addition to that, tidal disruption of planets in in eccentric orbit and circularization of the orbit due to tidal interactions with the host star can explain both upper and lower boundary of desert \citep{2018MNRAS.479.5012O}. TOI-181b is within this sparsely hot Neptune desert and is one of the largest planets discovered in this region; hence, it retains a substantial amount of its volatile envelope. This can be explained as, TOI-181b has enough gravitational binding energy to hold its massive H/He envelope against the incident hig energy X-ray or EUV radiation, which is already discussed in section \ref{p_env}.

\section{Conclusions}
\label{sec:5}
In this paper, we describe the validation of the sub-Saturn-like exoplanet TOI-181b. In order to validate, we obtained spectral information from the ESO science portal, which was collected by the 3.6-meter HARPS telescope, and transit photometry data from the MAST archive. We used two different false positive probability calculating software as an independen check to test the transit signals. The FPP for \texttt{VESPA} and \texttt{TRICERATOPS} was $1.68 \times 10^{-14}$ and $3.81 \times 10^{-4}$, respectively, which is significantly less than the 1 \% threshold. The high resolution image obtained with the Keck II/NIRC2 camera further suggests that the target is not being affected by any nearby companions. Additionally, we used \texttt{Juliet} to fit a transit and radial velocity model to the data. The calculated planetary parameters based on the best-fit model are: Orbital period 4.5320 $\pm$ 0.000002 days, Radius 0.63 $\pm$ 0.007 R$_J$, Transit Duration 0.083 $\pm$ 0.009 days, Equilibrium Temperature 895.10 $\pm$ 33.86 K, Insolation 102.43 $\pm$ 15.52 S$_{\earth}$, Mass 0.1452 $\pm$ 0.02 M$_J$, Density 0.75 $\pm$ 0.1 g cm$^{-3}$, and surface gravity 9.36 $\pm$ 1.35 m s$^{-2}$. We also find out the quadratic limb darkening parameters of the TESS infrared band and those are 0.6888 $\pm$ 0.13 and -0.2350 $\pm$ 0.21. Based on these calculated parameters, TOI-181b is characterised as a Sub-Saturn like planet. 

\section*{Acknowledgements}
This work has made use of the materials provided by Planet Hunters Coffee Chat, which is funded through NASA. This work made use of a virtual machine provided and maintained by Kevin Hardegree-Ullman to run the \texttt{VESPA} code. This research has made use of the Exoplanet Follow-up Observation Program website, which is operated by the California Institute of Technology, under contract with the National Aeronautics and Space Administration under the Exoplanet Exploration Program. We also thank the people who came forward and shared their experiences throughout the way. The following software were used in this research: \texttt{TLS} \citep{2019A&A...623A..39H}, \texttt{Lightkurve} \citep{2018ascl.soft12013L}, \texttt{Juliet} \citep{2019MNRAS.490.2262E}, \texttt{VESPA} \citep{2012ApJ...761....6M} and \texttt{TRICERATOPS} \citep{2020ascl.soft02004G}, \texttt{PLATYPOS} \citep{2022AN....34310105K}.

\section*{Data Availability}
The TESS photometry and the high angular resolution imaging data used in this article are available at the Mikulski Archive for Space Telescope (MAST) (\url{https://mast.stsci.edu/portal/Mashup/Clients/Mast/Portal.html}) and the ExoFOP-TESS website (\url{https://exofop.ipac.caltech.edu/tess/target.php?id=76923707}) respectively. Radial velocity data are provided in Table \ref{tab:01}. The data underlying this article will be made available on the reasonable request to the corresponding author.


\bibliographystyle{mnras}
\bibliography{Reference} 



\appendix

\section{Known Sample of Sub-Saturns}

\begin{table*}
    \centering
    \begin{tabular}{c c c c c c c c c c}
    \hline
       Planet & $N_P$ & $N_S$ & Radius & Mass & Density & Temperature & Eccentricity & [Fe/H] & Reference \\
              &      &         & $R_{\earth}$ & $M_{\earth}$ & g $cm^{-3}$ & K & & dex & \\
    \hline
    \hline
    CoRoT-8 b    & 1 & 1 & 6.39$^{+0.22} _{-0.22}$ & 69.29$^{+10.49} _{-13.03}$ & 1.45$^{+0.29} _{-0.29}$ & 844 & --- & 0.30 & \citet{borde2010transiting} \\
    GJ 436 b     & 1 & 1 & 4.09$^{+0.20} _{-0.20}$ & 23.42$^{+1.72} _{-1.72}$ & 2.01$^{+0.25} _{-0.25}$ & 669 & 0.14 & -0.03 & \citet{2004ApJ...617..580B} \\
    HAT-P-11 b   & 1 & 1 & 4.45$^{+0.10} _{-0.10}$ & 26.70$^{+2.22} _{-2.22}$ & 1.68$^{+0.16} _{-0.16}$ & 838 & 0.20 & 0.31 & \citet{2010ApJ...710.1724B} \\
    HAT-P-26 b   & 1 & 1 & 6.33$^{+0.81} _{-0.36}$ & 18.75$^{+2.23} _{-2.23}$ & 0.4$^{+0.10} _{-0.10}$ & 1001 & 0.12 & -0.04 & \citet{2011ApJ...728..138H} \\
    HATS-7 b     & 1 & 1 & 6.31$^{+0.52} _{-0.38}$ & 39.09$^{+4.45} _{-4.45}$ & 0.88$^{+0.24} _{-0.19}$ & 1070 & --- & 0.25 & \citet{2015ApJ...813..111B} \\
    K2-108 b     & 1 & 1 & 5.28$^{+0.54} _{-0.54}$ & 59.40$^{+4.40} _{-4.40}$ & 2.22$^{+0.77} _{-0.55}$ & 1446 & 0.18 & 0.33 & \citet{2017AJ....153..142P} \\
    K2-24 b      & 2 & 1 & 5.40$^{+0.20} _{-0.20}$ & 19.00$^{+2.20} _{-2.10}$ & 0.64$^{+0.12} _{-0.10}$ & 766 & --- & 0.42 & \citet{2016ApJ...818...36P} \\
    K2-24 c      & 2 & 1 & 7.50$^{+0.30} _{-0.30}$ & 15.40$^{+1.90} _{-1.80}$ & 0.20$^{+0.04} _{-0.03}$ & 605 & --- & 0.42 & 
    \citet{2016ApJ...818...36P}\\
    K2-27 b      & 1 & 3 & 4.45$^{+0.33} _{-0.33}$ & 29.10$^{+7.50} _{-7.40}$ & 1.80$^{+0.70} _{-0.55}$ & 910 & 0.25 & 0.13 & \citet{2017AJ....153..142P} \\
    K2-32 b      & 3 & 2 & 5.13$^{+0.28} _{-0.28}$ & 16.50$^{+2.70} _{-2.70}$ & 0.67$^{+0.16} _{-0.16}$ & 817 & --- & -0.02 & \citet{2017AJ....153..142P} \\
    K2-39 b      & 1 & 1 & 5.71$^{+0.63} _{-0.63}$ & 39.80$^{+4.40} _{-4.40}$ & 1.17$^{+0.47} _{-0.32}$ & 1670 & 0.15 & 0.43 & \citet{2017AJ....153..142P} \\
    Kepler-101 b & 2 & 1 & 5.77$^{+0.85} _{-0.79}$ & 51.10$^{+5.10} _{-4.70}$ & 1.45$^{+0.83} _{-0.48}$ & 1513 & 0.09 & 0.33 & \citet{2014ApJ...784...45R} \\
    Kepler-11 e  & 6 & 1 & 4.00$^{+0.20} _{-0.30}$ &  6.70$^{+1.20} _{-1.00}$ & 0.60$^{+0.20} _{-0.10}$ & 630 & 0.01 & -0.04 & \citet{2011Natur.470...53L} \\
    Kepler-18 c  & 3 & 1 & 5.22$^{+0.36} _{-0.34}$ & 18.40$^{+2.70} _{-1.90}$ & 0.57$^{+0.12} _{-0.10}$ & 979 & --- & 0.19 & \citet{2011ApJS..197....7C} \\
    Kepler-18 d  & 3 & 1 & 6.00$^{+0.40} _{-0.40}$ & 14.90$^{+1.80} _{-4.20}$ & 0.30$^{+0.10} _{-0.10}$ & 784 & --- & 0.19 & \citet{2011ApJS..197....7C} \\
    Kepler-223 d & 4 & 1 & 5.24$^{+0.26} _{-0.45}$ &  8.00$^{+1.50} _{-1.30}$ & 0.31$^{+0.12} _{-0.07}$ & 1040 & 0.04 & 0.06 & \citet{2014ApJ...784...45R} \\
    Kepler-223 e & 4 & 1 & 4.60$^{+0.27} _{-0.41}$ &  4.80$^{+1.40} _{-1.20}$ & 0.28$^{+0.12} _{-0.08}$ & 944 & 0.05 & 0.06 & \citet{2014ApJ...784...45R} \\
    Kepler-25 c  & 3 & 2 & 4.50$^{+0.08} _{-0.08}$ & 14.30$^{+2.70} _{-2.50}$ & 0.96$^{+0.24} _{-0.23}$ & 1080 & --- & -0.04 & \citet{2012MNRAS.421.2342S} \\
    Kepler-4 b   & 1 & 1 & 4.00$^{+0.21} _{-0.21}$ & 24.47$^{+3.81} _{-3.81}$ & 1.91$^{+0.36} _{-0.47}$ & 1650 & ---  & 0.17 & \citet{2010ApJ...713L.126B} \\
    Kepler-413 b & 1 & 2 & 4.35$^{+0.10} _{-0.10}$ & 67.00$^{+22.0} _{-21.0}$ & 3.20$^{+1.00} _{-1.00}$ & 348 & 0.12 & -0.20 & \citet{2014ApJ...784...14K} \\
    Kepler-56 b  & 3 & 1 & 5.10$^{+0.50} _{-0.70}$ & 32.00$^{+11.20} _{-0.50}$ & 1.40$^{+0.50} _{-0.50}$ & 1479 & --- & 0.37 & \citet{2013MNRAS.428.1077S} \\
    Kepler-79 d  & 4 & 1 & 6.91$^{+0.01} _{-0.01}$ & 11.30$^{+2.20} _{-2.20}$ & 0.09$^{+0.03} _{-0.03}$ & 626 & 0.03 & -0.02 & 
    \citet{2014ApJ...784...45R}\\
    Kepler-87 c  & 2 & 1 & 6.14$^{+0.29} _{-0.29}$ &  6.40$^{+0.80} _{-0.80}$ & 0.15$^{+0.02} _{-0.02}$ & 403 & 0.04 & -0.17 & \citet{ofir2014independent} \\
    \hline
    \end{tabular}
    \caption{Sample of sub-Saturns from \citet{2017AJ....153..142P}. $N_P$: Number of Planets, $N_S$: Number of Stars.}
    \label{tab:A1}
\end{table*}

\begin{table*}
    \centering
    \begin{tabular}{c c c c c c c c c c}
    \hline
       Planet & $N_P $ & $N_S$ & Radius & Mass & Density & Temperature & Eccentricity & [Fe/H] & Reference \\
              &      &         & $R_{\earth}$ & $M_{\earth}$ & g $cm^{-3}$ & K & & dex & \\
    \hline
    \hline
    GJ-3470 b     & 1 & 1 & 4.57$^{+0.18} _{-0.18}$ & 13.90$^{+1.50} _{-1.50}$ & 0.80$^{+0.13} _{-0.13}$ & 594 & --- & 0.20 & \citet{bonfils2012hot} \\
    HD 219666 b   & 1 & 1 & 4.71$^{+0.17} _{-0.17}$ & 16.60$^{+1.30} _{-1.30}$ & 0.87$^{+0.12} _{-0.11}$ & 1073 & 0.00 & 0.04 & \citet{esposito2019hd} \\
    HD 89345 b    & 1 & 1 & 6.86$^{+0.14} _{-0.14}$ & 35.70$^{+3.30} _{-3.30}$ & 0.61$^{+0.07} _{-0.07}$ & 1053 & 0.20 & 0.45 & \citet{2018MNRAS.478.4866V} \\
    K2-280 b      & 1 & 1 & 7.50$^{+0.44} _{-0.44}$ & 37.10$^{+5.60} _{-5.60}$ & 0.48$^{+0.13} _{-0.10}$ & 787 & 0.35 & 0.33 & \citet{2018AJ....156..277L} \\
    K2-60 b       & 1 & 1 & 7.66$^{+0.42} _{-0.42}$ & 135.4$^{+11.76} _{-11.76}$ & 1.7$^{+0.30} _{-0.30}$ & 1400 & 0.00 & 0.01 & \citet{2016ApJS..226....7C} \\
    Kepler-1656 b & 2 & 1 & 5.02$^{+0.53} _{-0.53}$ & 48.60$^{+4.20} _{-3.80}$ & 2.13$^{+0.87} _{-0.57}$ & 651 & 0.84 & 0.19 & \citet{2018AJ....156..147B} \\
    WASP-156 b    & 1 & 1 & 5.72$^{+0.22} _{-0.22}$ & 40.68$^{+3.18} _{-2.86}$ & 1.30$^{+0.10} _{-0.10}$ & 970 & 0.01 & 0.24 & \citet{demangeon2018discovery} \\
    \hline
    \end{tabular}
    \caption{Extension of previous sample by \citet{2020MNRAS.497.4423N}.}
    \label{tab:A2}
\end{table*}

\begin{table*}
    \centering
    \begin{tabular}{c c c c c c c c c c}
    \hline
       Planet & $N_P $ & $N_S$ & Radius & Mass & Density & Temperature & Eccentricity & [Fe/H] & Reference \\
              &      &         & $R_{\earth}$ & $M_{\earth}$ & g $cm^{-3}$ & K & & dex & \\
    \hline
    \hline
    AU Mic b    & 2 & 1 & 4.07$^{+0.17} _{-0.17}$ & 17.00$^{+5.00} _{-5.00}$ & 1.40$^{+0.40} _{-0.40}$ & 593 & 0.00 & ---  & \citet{2020Natur.582..497P} \\
    HATS-37 A b & 1 & 2 & 6.79$^{+0.18} _{-0.18}$ & 31.47$^{+13.35} _{-13.35}$ & 0.55$^{+0.24} _{-0.24}$ & 1085 & 0.00 & 0.05 & \citet{2020AJ....160..222J} \\
    HATS-38 b   & 1 & 1 & 6.88$^{+0.19} _{-0.19}$ & 23.52$^{+3.50} _{-3.50}$ & 0.40$^{+0.07} _{-0.07}$ & 1294 & 0.00 & -0.10 & \citet{2020AJ....160..222J} \\
    K2-98 b     & 1 & 1 & 4.30$^{+0.30} _{-0.20}$ & 32.20$^{+8.10} _{-8.10}$ & 2.15$^{+0.67} _{-0.60}$ & 1102 & 0.00 & -0.20 & \citet{2016AJ....152..193B} \\
    KOI-1783.02 & 2 & 1 & 5.44$^{+0.52} _{-0.30}$ & 15.00$^{+4.30} _{-3.60}$ & 0.51$^{+0.21} _{-0.15}$ & --- & 0.00 & 0.11 & \citet{2020AJ....159..108V} \\
    LP 714-47 b & 1 & 1 & 4.70$^{+0.30} _{-0.30}$ & 30.80$^{+1.50} _{-1.50}$ & 1.70$^{+0.30} _{-0.30}$ & 700 & 0.04 & 0.41 & \citet{dreizler2020carmenes} \\
    LTT 9779 b  & 1 & 1 & 4.72$^{+0.23} _{-0.23}$ & 29.32$^{+0.78} _{-0.81}$ & 1.54$^{+0.12} _{-0.12}$ & 1978 & 0.00 & 0.27 & \citet{2020NatAs...4.1148J} \\
    NGTS-14 A b & 1 & 2 & 4.98$^{+0.34} _{-0.34}$ & 29.24$^{+3.81} _{-3.81}$ & 1.40$^{+0.33} _{-0.33}$ & 1143 & 0.00 & 0.10 & \citet{smith2021ngts} \\
    TOI-1710 b  & 1 & 1 & 5.34$^{+0.11} _{-0.11}$ & 28.30$^{+4.70} _{-4.70}$ & 0.94$^{+0.22} _{-0.22}$ & 687 & 0.16 & 0.10 & \citet{konig2022warm} \\
    TOI-1728 b  & 1 & 1 & 5.05$^{+0.16} _{-0.17}$ & 26.78$^{+5.43} _{-5.13}$ & 1.14$^{+0.26} _{-0.24}$ & 767 & 0.06 & 0.25 & \citet{2020ApJ...899...29K} \\
    TOI-181 b   & 1 & 1 & 6.96$^{+0.09} _{-0.10}$ & 46.17$^{+2.71} _{-7.83}$ & 0.76$^{+0.02} _{-0.10}$ & 895 & 0.15 & 0.27 & This Work \\
    TOI-257 b   & 1 & 1 & 7.16$^{+0.15} _{-0.15}$ & 43.86$^{+7.31} _{-7.31}$ & 0.65$^{+0.12} _{-0.11}$ & 1027 & 0.00 & 0.17 & \citet{addison2021toi} \\
    TOI-3884 b  & 1 & 1 & 6.00$^{+0.18} _{-0.18}$ & 16.50$^{+3.50} _{-1.80}$ & 0.42$^{+0.10} _{-0.06}$ & 463 & --- & 0.23 & \citet{almenara2022toi} \\
    TOI-421 c   & 2 & 2 & 5.09$^{+0.16} _{-0.15}$ & 16.42$^{+1.06} _{-1.04}$ & 0.69$^{+0.08} _{-0.07}$ & 674 & 0.15 & -0.02 & \citet{2020AJ....160..114C} \\
    TOI-532 b   & 1 & 1 & 5.82$^{+0.19} _{-0.19}$ & 61.50$^{+9.70} _{-9.30}$ & 1.72$^{+0.31} _{-0.31}$ & 867 & 0.00 & 0.34 & \citet{2021AJ....162..135K} \\
    TOI-674 b   & 1 & 1 & 5.25$^{+0.17} _{-0.17}$ & 23.60$^{+3.30} _{-3.30}$ & 0.91$^{+0.15} _{-0.15}$ & 635 & 0.00 & 0.17 & \citet{murgas2021toi} \\
    WASP-166 b  & 1 & 1 & 7.06$^{+0.34} _{-0.34}$ & 32.10$^{+1.59} _{-1.59}$ & 0.54$^{+0.09} _{-0.09}$ & 1270 & 0.00 & 0.19 & \citet{2019MNRAS.488.3067H} \\
    \hline
    \end{tabular}
    \caption{Extension of previous list by this paper. }
    \label{tab:A3}
\end{table*}

\bsp	
\label{lastpage}
\end{document}